\documentclass{article}

\usepackage{PRIMEarxiv}

\usepackage[utf8]{inputenc} 
\usepackage[T1]{fontenc}    
\usepackage{hyperref}       
\usepackage{url}            
\usepackage{booktabs}       
\usepackage{amsfonts}       
\usepackage{nicefrac}       
\usepackage{microtype}      
\usepackage{fancyhdr}       
\usepackage{graphicx}       
\usepackage{natbib}
\usepackage{amsmath}
\usepackage{placeins}
\usepackage{soul}
\usepackage{xcolor}
\usepackage{listings}

\definecolor{codegreen}{rgb}{0,0.6,0}
\definecolor{codegray}{rgb}{0.4,0.4,0.4}
\definecolor{codepurple}{rgb}{0.58,0,0.82}
\definecolor{backcolour}{rgb}{0.99,0.99,0.99}
\definecolor{codecomment}{rgb}{0.4,0.4,0.4}
\lstdefinestyle{mystyle}{
	backgroundcolor=\color{backcolour},   
	commentstyle=\color{codecomment},
	keywordstyle=\color{codepurple},
	numberstyle=\tiny\color{gray},
	stringstyle=\color{codegreen}, 
    columns=fullflexible,
	breakatwhitespace=false, 
	escapeinside={\%}{)},     
	breaklines=true,                 
	captionpos=b,                    
	keepspaces=true,                 
	numbers=left,                    
	numbersep=1pt,                  
	showspaces=false,                
	showstringspaces=false,
	showtabs=false,                  
	tabsize=2
}
\graphicspath{{media/}}     

\let\oldtabular=\tabular
\def\tabular{\large\oldtabular}
  
\title{ \textcolor{black}{Enhanced Level-Set Method for free surface flow applications} }

\author{
  Paulin FERRO, Paul LANDEL, Carla LANDRODIE, Simon GUILLOT, Marc PESCHEUX  \\
  SARL G-MET Technologies \\
  62 rue d'Hyères, 83140 Six-Fours-Les-Plages\\
}

\begin{document}
\maketitle

\begin{abstract}
This publication presents a solver using the Level-Set method (\cite{SUSSMAN1994146}) for incompressible two phase flows with surface tension. A one fluid approach is adopted where both phases share the same velocity and pressure field. The Ghost Fluid Method (\cite{Fedwick1999}) is also used. An efficient and pragmatic solution is proposed to avoid interface displacement during the reinitialization of the Level-Set field. A solver called \textit{LSFoam} is created in the OpenFOAM (\cite{Weller1998}) framework with a consistent Rhie \& Chow interpolation (\cite{Cubero2007}). This solver is tested on several test cases, covering different scales and flow configurations: rising bubble test case, \citet{Hysing2007}, Rayleigh-Taylor instability simulations \citet{Puckett1997}, Ogee spillway flow \citet{erpicum2018experimental}, 3D dambreak simulation with a square cylinder obstacle \citet{GomezGesteira2013} and KVLCC2 steady resistance calculations \citet{Larsson2014}. Overall results are in excellent agreement with reference data and the present approach is very promising for moderate free surface deformations.
\end{abstract}


\section{Introduction}


\textcolor{black}{When simulating free surface flows, both the Volume-of-Fluid method (VoF, \citet{Hirt1981}) and the Level-Set method (LS, \citet{SUSSMAN1994146}) can be used to capture interface between two phases. The Volume of Fluid (VoF) method is widely adopted in industry for multiphase flow simulations and is integrated into major commercial software such as Star-CCM+ (\cite{starCCMdoc}), Fluent (\cite{fluent2017ansys}), and OpenFOAM (\textit{interFoam} solver). In this work, the term "industrial" refers to a challenging context of various 3D flow configurations with different cell types, high maximal non-orthogonality (above 80°) and maximal Courant number higher than 50. The VoF method popularity stems from its inherent ability to conserve mass, making it a natural choice within industrial constraints. In contrast, the original Level-Set (LS) method, based on the function $\psi$ being the signed distance to the interface, has limited dissemination to commercial software and industrial applications. It is likely due to its lack of mass conservation property. However, the LS method offers numerical advantages such as more accurate and reliable curvature computations as well as a user controlled interface thickness.} Additionally, with the LS approach, the distance to the interface is readily available, making it easy to combine with the Ghost Fluid Method (GFM), \citet{Fedwick1999}. To capture the interface motion, the LS function is advected in a velocity field. However, when the tangential component of the normal velocity gradient is non-zero, such a procedure will break the distance property of the Level-Set function ($\left|\boldsymbol{\nabla}\psi\right|\neq1$). This behavior will generate unacceptable mass variations. Hence, a reinitialization procedure, \citet{SUSSMAN1994146}, is often used for recovering the signed distance property of the LS function.
To limit mass variations, researchers traditionally use a combination of 5th order WENO spatial schemes \citet{Liu1994}, and Runge Kutta (RK2 or RK3) temporal scheme for the resolution of both transport and reinitialization equations (\citet{Huang2007}, \citet{Lalanne2015}). Even with this numerical treatment, there is still no guarantee that mass will be conserved, \citet{Gu2018}. Moreover, even if WENO schemes have been promisingly used for unstructured grids (\citet{Martin2018}, \citet{Gaertner2020}), they remain complex to use for industrial applications. Indeed, WENO schemes require low Courant numbers ($<1/3$ in 3D, \citet{Titarev2004}) which can be very restricting to decrease the calculation time when only the steady state regime is of interest (typically for ship resistance assessment). The sub-cell fix method, \cite{Hartmann2010}, \cite{Russo2000}, has been developed to limit interface displacements during the reinitilization, but its application is limited to Cartesian grids. Moreover, \cite{Sun2010} have shown that the zero Level-Set can be disturbed even with the sub-cell fix method. Applications of the original approach of \citet{SUSSMAN1994146} for 3D industrial situations remain limited, \cite{Khosronejad2019}, \cite{Park2005}, \cite{Huang2007}, and restricted to structured grids. Hence, the approach efficiency, accuracy, robustness and range of applications for such situations remain unclear. Therefore, using the available literature, this work aims to develop an efficient Level-Set/GFM solver for arbitrary polyhedral cells, incorporating enhancements to the the Level-Set method while maintaining the simplicity of the original approach of \citet{SUSSMAN1994146}. The solver, deployed within 2nd order finite volume OpenFOAM framework \citet{Weller1998} and referred as \textit{LSFoam}, is tested for various flow configurations:


\begin{itemize}
	\item Rising bubble (\citet{Hysing2007}),
	\item Rayleigh-Taylor instability simulations (\citet{Puckett1997}),
    \item Ogee spillway flow (\citet{erpicum2018experimental}),
	\item 3D dambreak simulation (\citet{GomezGesteira2013}),
	\item KVLCC2 steady resistance calculations (\citet{Larsson2014})	
\end{itemize}

\section{Mathematical and numerical procedure}

\subsection{The Level-Set equation}\label{LS}

The computational domain is divided into two pieces $\varOmega^{+}$
and $\varOmega^{-}$ separated by the interface $\Gamma$. Where $\varOmega^{+}$
and $\varOmega^{-}$ represent respectively the domain of heavy and
light phases. For a given point $\boldsymbol{x}$, the Level-Set function
$\psi(x)$ is defined by the shortest distance $d$ to the interface.
It is signed depending on the point domains (\ref{eq:psi}),
so that both numerical transport and stability near the interface
are improved by avoiding $\boldsymbol{\nabla}\psi$ discontinuities.

\begin{equation}
    \psi(\boldsymbol{x})=\begin{cases}
        0 & \boldsymbol{x}\in\Gamma\\
        d(\boldsymbol{x}) & \boldsymbol{x}\in\varOmega^{+}\\
        -d(\boldsymbol{x}) & \boldsymbol{x}\in\varOmega^{-}
    \end{cases}\label{eq:psi}
\end{equation}

A consequence of this definition is that $\forall x , \left|\boldsymbol{\nabla}\psi\right|=1$.

\begin{figure}[h!]
        \centering
        \includegraphics[scale = 0.4]{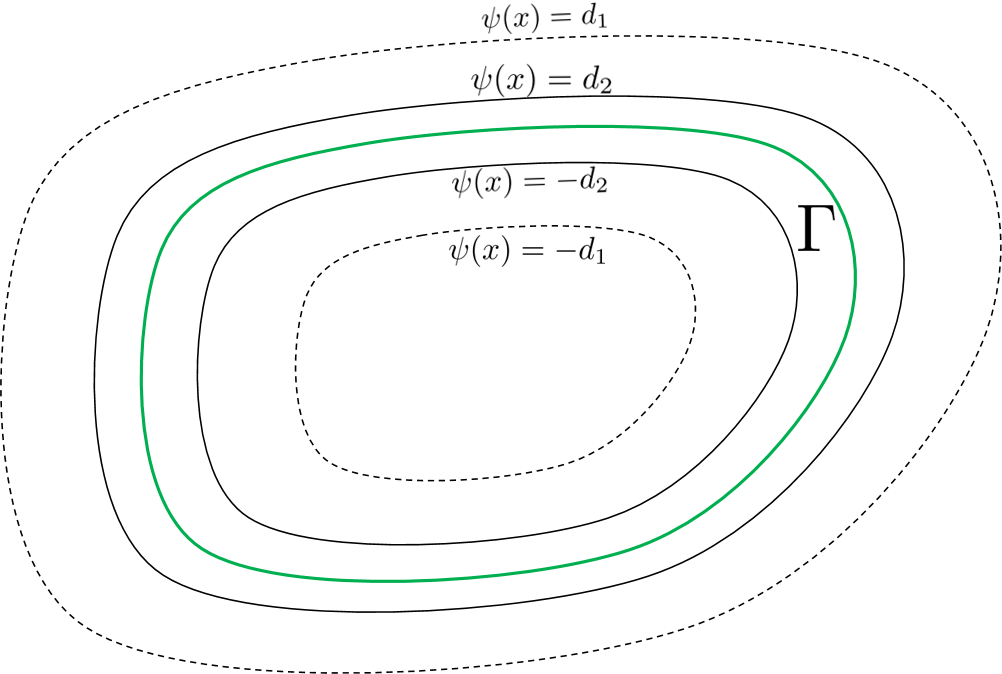}
        \caption{\centering Illustration of the definition of the Level-Set function. Each iso-contour $\left\{ \boldsymbol{x}\mid\psi(\boldsymbol{x})=d\right\} $ define all the points $\boldsymbol{x}$ at a distance $d$ from the interface $\Gamma$.
        } \label{fig1}           
\end{figure}

Interface normal $\boldsymbol{n}$ and curvature $\kappa$ can be
computed as follow:

	\begin{equation}
		\boldsymbol{n}=\frac{\boldsymbol{\nabla}\psi}{\left|\boldsymbol{\nabla}\psi\right|}\label{eq:-2}
	\end{equation}

	\begin{equation}
		\kappa=\boldsymbol{\nabla}\cdot\boldsymbol{n}\label{eq:-3}
	\end{equation}

	The Level-Set function $\psi$ is advected in a flow field $\boldsymbol{u}$ (assumed incompressible) using a transport equation:

	\begin{equation}
		\frac{\partial\psi}{\partial t}+\boldsymbol{\nabla}\cdot(\boldsymbol{u}\psi)=0\label{eq:}
	\end{equation}

	The resolution of the transport equation is done with a sub-cycling strategy,  \cite{Leroyer2011}. The time step is subdivided into $N$ sub-cycles enhancing  mass conservation. However, the transport of the Level-Set function will breach the distance property and is likely to cause mass variations. To limit
	such consequences, a reinitialization procedure is adopted by solving
	the following (eikonal) equation (\citet{SUSSMAN1994146}):

	\begin{equation}
		\frac{\partial\psi}{\partial\tau}=S\left(\psi_{0}\right)\left(1-\left|\boldsymbol{\nabla}\psi\right|\right)\label{eq:-1}
	\end{equation}

	Where $S$ is the sign function, $\psi_{0}$ the initial
	value of the Level-Set function and $\tau$ has the dimension
	of a length. The function $S$ is smoothed to improve the numerical
	resolution in the vicinity of the interface using the following expression
	\citet{books/lib/OsherF03}:

	\begin{equation}
		S\left(\psi_{0}\right)=\frac{\psi_{0}}{\sqrt{\psi_{0}^{2}+\left|\boldsymbol{\nabla}\psi_{0}\right|\epsilon^{2}}}\label{eq:-1-1}
	\end{equation}

	The term $\left|\boldsymbol{\nabla}\psi_{0}\right|$ helps the reinitialization process when the initial field $\psi_{0}$ is away from the current Level-Set field. $\epsilon$ can be a constant chosen as 2 or 3 the size of a user-defined reference cell. However, this can be prejudicial with non-uniform grids. To tackle this issue, $\epsilon(x)$ becomes a variable computed on the domain for each cell $C(x)$: 
	
	\begin{equation}
		\epsilon ( \boldsymbol{x} ) = k l_e
		\label{eq:espilon}
	\end{equation}

with $k$ a user defined constant (usually $k=2$) and $l_e(x)$ the length of the cell edge that have the highest scalar product with the interface normal vector so that: $$\boldsymbol{l_e} \cdot \boldsymbol{n} =  max(\boldsymbol{l_{e_{i}}} \cdot \boldsymbol{n}), \: \forall e_i \in C(x)$$
	
	\citet{Kim2021} used an Euler explicit resolution of Equation \ref{eq:-1}
	based on the code of \citet{Yamamoto2017}. However, the explicit
	method has a limited range of applicability, especially for unstructured
	grids. Instead, \citet{Sussman1998} defined:
	
	\begin{equation}
		\boldsymbol{w}=S\left(\psi_{0}\right)\boldsymbol{n}
    \end{equation}		
			
	which leads	to the following form of the reinitialization equation:

	\begin{equation}
		\frac{\partial\psi}{\partial\tau}+\boldsymbol{\nabla}\cdot(\boldsymbol{w}\psi)-\psi \boldsymbol{\nabla}\cdot\boldsymbol{w}=S\left(\psi_{0}\right)\label{eq:-1-2}
	\end{equation}

Equation \ref{eq:-1-2} is more suitable for implicit finite volume discretization \cite{Vukcevic2014}, easing the use of unstructured meshes and improving the numerical stability. In this study, the reinitialization equation \ref{eq:-1-2} is implicitly solved with the first order Euler scheme. The convective term $\boldsymbol{\nabla}\cdot(\boldsymbol{w}\psi)$ is discretized using 2nd order TVD MUSCL scheme, \citet{vanLeer1979}, and the third term of Equation \ref{eq:-1-2} is treated implicitly. \textcolor{black}{The reinitialization equation is solved for a user defined  number of iteration with $i$ is the iteration counter. For a given cell $P$ with volume $V_P$, surrounded with neighboring cells $N_f$ that share the face $f$ of cell $P$, the discretization of Equation \ref{eq:-1-2} can be expressed using Gauss's theorem for divergence operators as follows:} 
 \textcolor{black}{
 	\begin{equation}
		\frac{\psi^i_P-\psi_P^{i-1}}{\tau} V_P+ \underset{f}{\sum} [\boldsymbol{w}^{i-1}]_f\cdot \boldsymbol{S}_f \psi_f^{i}-\psi_P^{i} \underset{f}{\sum}[\boldsymbol{w}^{i-1}]_f\cdot \boldsymbol{S}_f=\frac{\psi_{P}^0}{\sqrt{{\psi_P^0}^{2}+\left|\boldsymbol{\nabla}\psi_P^{0}\right|\epsilon_P^{2}}}V_P
        \label{eq:FVMreini}
	\end{equation}}
 
\textcolor{black}{Where, $\psi^i_P$ represents the unknown cell-centered values, and $\psi_P^{0}$ is the signed distance function before the reinitialization procedure ($i=0$) at cell $P$. The bracket $[.]_f$ indicates a linear interpolation from cell center to face center. For the convective term, $\psi_f^{i}$ is written as a linear function of $\psi_P^{i}$ and $\psi_{N_f}^{i}$ using MUSCL scheme, \citet{vanLeer1979}. The non-linearity in the divergence terms are handled by using previous values of ${w}$ at iteration $i-1$}. Authors generally use meshes with uniform cell size and define $\tau=0.5\Delta z$ where $\Delta z$ is the grid size. Nevertheless, for unstructured, \textcolor{black}{\st{highly}} distorted, or non-uniform meshes commonly used in industrial situations, this approach is not optimal as the cell size can vary significantly, from small near the walls to large in the far-field region. We define the reinitialization Courant number $\gamma$ \textcolor{black}{by mimicking the definition used for momentum equation}:
	
	\begin{equation}
		\gamma(\boldsymbol{x})=  \frac{\tau}{V\left(\boldsymbol{x}\right)}\underset{f}{\sum} \boldsymbol{w}_f\textcolor{black}{\cdot \boldsymbol{S}_f}\label{eq:gammaMax}
	\end{equation}
	
Where $V\left(\boldsymbol{x}\right)$ is the volume cell size, \textcolor{black}{ $\boldsymbol{w}_f$ the linear interpolation of $\boldsymbol{w}$ from cell to face} and $\underset{f}{\sum}$ the \textcolor{black}{sum over cell faces \st{face average operator}}. In practice, the magnitude of $\boldsymbol{w}$ is not always equal to 1, especially before reinitialization, due to numerical errors or when $\psi$ deviates significantly from the signed distance function. In this study, Equation \ref{eq:-1-2} is solved with a local time stepping approach (LTS) where the spatial time step $\tau$ is manipulated locally based on a given maximal reinitialization Courant number $\gamma_{max}$:
	
	\begin{equation}
		\frac{1}{\tau\left(\boldsymbol{x}\right)} = max\left(\frac{1}{\Delta z\left(\boldsymbol{x}\right)},\frac{\underset{f}{\sum}\boldsymbol{w}_f \textcolor{black}{\cdot \boldsymbol{S}_f} }{\gamma_{max}V\left(\boldsymbol{x}\right)}\right)\label{eq:tauVariable}
	\end{equation}
	
	The above procedure	allows to maximize the spatial time step for each cell based on $\gamma_{max}$ value. The spatial time step is limited by the local cell size $\Delta z\left(\boldsymbol{x}\right)$. In practice, interface displacements can occur during the reinitialization procedure leading to mass variations. If the reinitialization frequency and/or the number of iterations is too high, significant interface displacements will occur. Traditionally, authors solve the eikonal equation for a given reinitialization frequency and for a number of iterations \cite{henriPhD}, \cite{Johansson2011ImplementationOA}. Hence, having to determine the optimal iteration number and frequency is the major drawback of this method. Moreover, these parameters are strongly case-dependent, which is limiting its use for a wide range of complex situations \textcolor{black}{encountered under industrial context}. 
	
\begin{figure}[h!]
        \centering
        \includegraphics[scale=0.25]{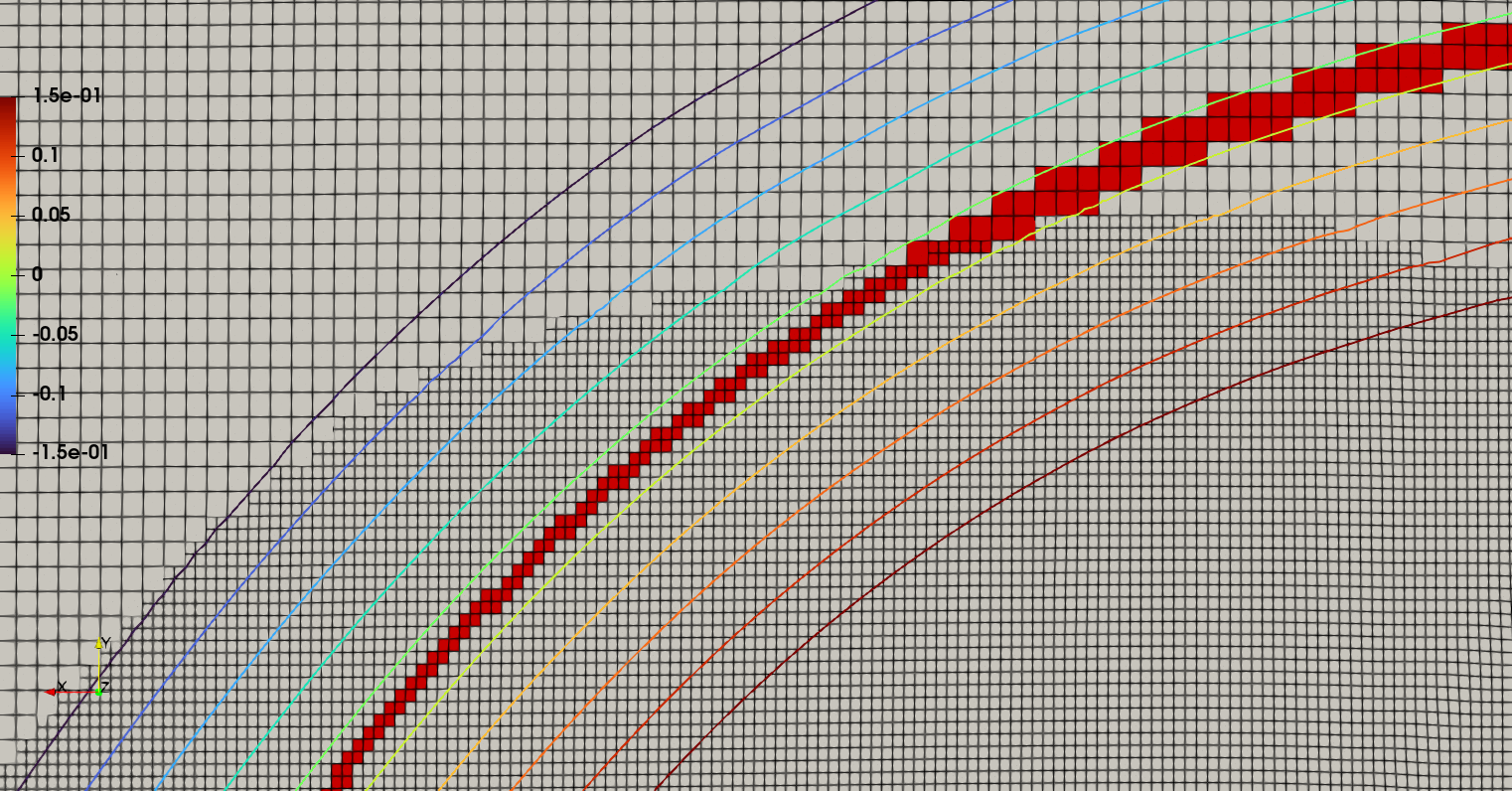}
        \caption{\centering Illustration of the anchoring cells (coloured in red) used during the reinitialization procedure - Level-Set contours are also plotted after reinitialization.}
        \label{fig:levelSetFixePoint}
    \end{figure}

	To tackle all this, a simple and pragmatic solution that avoids any interface displacements during the resolution of the eikonal equation is proposed. The set of cells (called anchoring cells) crossing the zero Level-Set contour are detected. \textcolor{black}{The simplified algorithmic procedure is presented in algorithm \ref{lst:anchoring_cell}. The detection is done by looping over the mesh faces and by marking all the cells that share the faces satisfying the criterion $\psi_{P}\psi_{N}<0$ (Equation \ref{eq:-10}), where $P$ and $N$ the face owner and neighboring cells. The anchoring cells are distinguished using a boolean list.} The corresponding Level-Set values are stored before the reinitialization procedure. Then, after each iteration of the eikonal equation resolution, the Level-Set values of the anchoring cells are restored so that the zero Level-Set contour remains undisturbed during the reinitialization. \textcolor{black}{The efficiency of the present approach is demonstrated in chapter \ref{LS_testcase1}.} This method is consistent with the GFM, that uses the same criterion to identify interface faces. The present approach is illustrated in Figure \ref{fig:levelSetFixePoint}, where the anchoring cells are colored in red. Level-Set contours are also plotted after reinitialization.


    \newpage

    \begin{lstlisting}[caption={redistanciation procedure simplified algorithm},label={lst:anchoring_cell},language=Python]
      #Starting reinitialisation
      store %$\psi_0$)
      compute %S($\psi_0$))     
      initialize list anchoringCell #False for all cells
    
      #Looping over all mesh faces
      for face in mesh: 
            if %$\psi_{P}\psi_{N}<0$): # P and N sharing face f
               anchoringCell[P] = True 
               anchoringCell[N] = True 
             
      while reinitialization iteration:
          solve eikonal equation
          restore %$\psi = \psi_0$) for anchoringCell = True              
    \end{lstlisting}   
 
	The Level-Set function is used to compute the phase fraction with an hyperbolic filtering function. The volume fraction $\alpha$ is calculated using the standard hyperbolic filtering function:

	\begin{equation}
		\alpha=\frac{1}{2}\left(tanh\left(\frac{\pi\psi}{\epsilon(\boldsymbol{x})}\right)+1\right)\label{eq:-15}
	\end{equation}

	The phase fraction $\alpha$ is then used to calculate the
	mixture viscosity $\mu$ as:

	\begin{equation}
		\mu=\alpha\mu^{+}+\left(1-\alpha\right)\mu^{-}
	\end{equation}

	\begin{equation}
		\begin{cases}
			\rho=\rho^{+} & if\;\psi>0\\
			\rho=\rho^{-} & if\;\psi<0
		\end{cases}
	\end{equation}

	Where $\mu^{+}$and $\mu^{-}$ are the dynamic viscosity
	of heavy and light phases and $\rho^{+}$ and $\rho^{-}$ their density.
    
	\subsection{Consistent pressure-velocity coupling}

	The momentum equation in each phase is given as follows, \cite{Trujillo2021}:

	\begin{equation}
    \textcolor{black}{\frac{\partial\boldsymbol{u}}{\partial t}+\boldsymbol{\nabla}\cdot(\boldsymbol{u}\otimes\boldsymbol{u})
    =\frac{1}{\rho}\left(-\boldsymbol{\nabla}p+\rho\boldsymbol{g}+\boldsymbol{\nabla}\cdot\left[\mu_{eff} (\nabla\boldsymbol{u} + (\nabla\boldsymbol{u})^T) - \mu_{eff}\frac{2}{3} \mathrm{tr}(\nabla  \boldsymbol{u})^T \right]
    +\sigma\kappa\boldsymbol{n}\delta_{\Gamma}\right)}\label{eq:-4}
	\end{equation}

	Where $\sigma$ is the surface tension coefficient, $\boldsymbol{n}$
	the interface normal (Equation \ref{eq:-2}), $\delta_{\Gamma}$ a
	surface Dirac function that equals one at the interface and zero elsewhere,
	and $\mu_{eff}$ the effective viscosity \textcolor{black}{(sum of the molecular $\mu$ and turbulent viscosity $\mu_t$)}. In Equation \ref{eq:-4},
	the pressure \textcolor{black}{$p$} is replaced by its decomposition into dynamic (or piezometric)
	$p_{d}$ and hydrostatic parts:

	\begin{equation}
		p=p_{d}+\rho\boldsymbol{g}\cdot\boldsymbol{x}
	\end{equation}

	\begin{equation}
		\boldsymbol{\nabla}p=\boldsymbol{\nabla}p_{d}+\boldsymbol{\nabla}\rho\boldsymbol{g}\cdot\boldsymbol{x}+\rho\boldsymbol{g}
	\end{equation}

\textcolor{black}{With $\boldsymbol{x}$ the coordinate vector}. The reason for using $p_{d}$ is to avoid any sudden changes
	of the pressure field at the boundaries for hydrostatic problems,
	\citet{Rusche2002}. Then, assuming a piece-wise density field, the
	momentum balance takes the following form in each phase domain $\varOmega^{+}$
	and $\varOmega^{-}$:

	\begin{equation}
		   \textcolor{black}{ \frac{\partial\boldsymbol{u}}{\partial t}+\boldsymbol{\nabla}\cdot(\boldsymbol{u}\otimes\boldsymbol{u})
    =\frac{1}{\rho}\left(\boldsymbol{\nabla}p_{d}+\boldsymbol{\nabla}\cdot\left[\mu_{eff} (\nabla\boldsymbol{u} + (\nabla\boldsymbol{u})^T) - \mu_{eff}\frac{2}{3} \mathrm{tr}(\nabla  \boldsymbol{u})^T \right]\right)},\;in\;\varOmega^{+}\;or\;\varOmega^{-}\;\label{eq:-4-1}
	\end{equation}

	Having a continuous velocity and dynamic viscosity fields,
	Equation \ref{eq:-4-1} is completed by the following set of jump
	conditions at the interface $\Gamma$:

	\begin{equation}
		\left[\frac{\boldsymbol{\nabla}p_{d}}{\rho}\right]_{\Gamma}=0\label{eq:-8}
	\end{equation}

	\begin{equation}
		\left[p_{d}\right]_{\Gamma}=\sigma\kappa+\left[\rho\right]_{\Gamma}\boldsymbol{g}\cdot\boldsymbol{x}_{\Gamma} \textcolor{black}{\equiv} \mathcal{H}\label{eq:-9}
	\end{equation}
	Where $\boldsymbol{x}_{\Gamma}$ is the interface coordinate vector.
	The bracket notation $\left[.\right]_{\Gamma}$ indicates a jump value
	between both sides of the interface. The momentum equation \ref{eq:-4-1}
	is discretized without the pressure gradient term. The remaining terms
	are treated implicitly, excepting the term $\frac{1}{\rho}\boldsymbol{\nabla}\cdot\left[\mu_{eff}(\nabla\textbf{u})^T - \mu_{eff}\frac{2}{3} \mathrm{tr}(\nabla  \textbf{u})^T \right]$.
	The semi-discretized momentum equation takes the following form:

	\begin{equation}
		a_{P}\boldsymbol{u}_{P}=-\frac{\boldsymbol{\nabla}p_{d}}{\rho}-\underset{f}{\sum}a_{N}\boldsymbol{u}_{N}+\boldsymbol{s}(\boldsymbol{u}_{P})=-\frac{\boldsymbol{\nabla}p_{d}}{\rho}+\boldsymbol{H}(\boldsymbol{u}_{P})\label{eq:-5}
	\end{equation}

	The coefficient $a_{P}$ and $a_{N}$ are respectively the
	diagonal and off-diagonal terms of the discretized momentum equations.
	Explicit contributions \textcolor{black}{$\boldsymbol{s}(\boldsymbol{u})$} are grouped in $\boldsymbol{H}(\boldsymbol{u})$.
	The summation stands for all the faces shared by the owner cell P
	and its neighboring cells N. To avoid checkerboard oscillations on collocated grids, the so-called \cite{Rhie1983} interpolation is used to obtain the face velocity by mimicking Equation \ref{eq:-5}. In order to avoid relaxation factor and time step dependencies, the interpolation needs to be done in a consistent way. The approach of \cite{Cubero2007} that has initially been developed for a single phase, is applied  here for two-phase flows. The coefficient $a_{P}$ is decomposed between its temporal $a_{t}$ and spatial $a_{s}$ parts. The old time contribution is taken out from $\boldsymbol{H}(\boldsymbol{u})$ (first order \textit{Euler} scheme here as an example) and relaxation is applied to Equation \ref{eq:-5}, resulting in:
	
	\begin{equation}
		(a_{t}+a_{s})\boldsymbol{u}_{P}=-\alpha_{u}\frac{\boldsymbol{\nabla}p_{d}}{\rho}+\alpha_{u}\boldsymbol{H}(\boldsymbol{u}_{P}) + \alpha_{u}a_{t}\boldsymbol{u}_{P}^{0}+(1-\alpha_{u})(a_{t}+a_{s})\boldsymbol{u}_{P}^{k-1}\label{eq:-16}
	\end{equation}
	Where $\alpha_{u}$ is the relaxation factor, $\boldsymbol{u}_{P}^{k-1}$ the velocity field of the previous non linear iteration (PIMPLE loop), and $\boldsymbol{u}_{P}^{0}$ the previous time step velocity field. In the case of 2nd order time discretization (backward scheme), the previous equation can be easily adapted by adding the old-old contribution $\boldsymbol{u}_{P}^{00}$ and the proper coefficients. The Equation \ref{eq:-16} is re-arranged as follows:
	
	\begin{equation}
		\boldsymbol{u}_{P}=\frac{1}{1+d}\left(-\frac{\alpha_{u}}{a_{s}}\frac{\boldsymbol{\nabla}p_{d}}{\rho}+\frac{\alpha_{u}}{a_{s}} \boldsymbol{H}(\boldsymbol{u}_{P}) + \alpha_{u}d\boldsymbol{u}_{P}^{0}\right)+(1-\alpha_{u})\boldsymbol{u}_{P}^{k-1} \label{eq:-20}
	\end{equation}
	Where $d = \frac{a_{t}}{a_{s}}$. The optional resolution of Equation \ref{eq:-20}, with an explicit pressure field, is the momentum predictor
	step. Following \cite{Rhie1983}, the face velocity equation is then obtained by mimicking Equation \ref{eq:-20} at faces (written in term of flux $\phi$ \textcolor{black}{[$m^3$/$s$]}). 
	
	\begin{equation}
		\phi_{f}=\frac{1}{1+[d]_{f}}\left(- \left[\frac{\alpha_{u}}{a_{s}}\right] _{f} \left(\frac{\boldsymbol{\nabla}p_{d}}{\rho}\right)_{f}+\left[\frac{\alpha_{u}\boldsymbol{H}(\boldsymbol{u}_{P}) }{a_{s}}\right] _{f} \textcolor{black}{\cdot \boldsymbol{S}_{f}} + \alpha_{u}[d]_{f}\phi_{f}^{0}\right)+(1-\alpha_{u})\phi_{f}^{k-1}\label{eq:-18}
	\end{equation}

    Where $[.]$ is the operator that linearly interpolates from cell center to face center. The continuity equation is then used on Equation \ref{eq:-18} to obtain the pressure Poisson equation in its finite volume discretized form:

	\begin{equation}
		\underset{f}{\sum}\frac{\left[\frac{\alpha_{u}}{a_{s}}\right] _{f} }{1+[d]_{f}}\left(\frac{\boldsymbol{\nabla}p_{d}}{\rho}\right)_{f} \cdot\boldsymbol{S}_{f}
		=
		\underset{f}{\sum}\frac{1}{1+[d]_{f}}\left( \left[\frac{\alpha_{u}\boldsymbol{H}(\boldsymbol{u}_{P})}{a_{s}}\right]_{f} \textcolor{black}{\cdot \boldsymbol{S}_{f}} + \alpha_{u}[d]_{f}\phi_{f}^{0}\right)+(1-\alpha_{u})\phi_{f}^{k-1} \label{eq:-7-1}
	\end{equation}

	The mesh non-orthogonality is handled using the over-relaxed approach, \citet{Jasak1996}. The surface vector $\boldsymbol{S}_{f}$ is then decomposed in two parts: the orthogonal $\boldsymbol{\delta}$ and non-orthogonal $\boldsymbol{k}$ contributions:

	\begin{equation}
		\left(\frac{\boldsymbol{\nabla}p_{d}}{\rho}\right)_{f}\cdot\boldsymbol{S}_{f}=\underset{implicit}{\underbrace{\left(\frac{\boldsymbol{\nabla}p_{d}}{\rho}\right)_{f}\cdot\boldsymbol{\delta}}}+\underset{explicit}{\underbrace{\left(\frac{\boldsymbol{\nabla}p_{d}}{\rho}\right)_{f}\cdot\boldsymbol{k}}}\label{eq:-7-1-2}
	\end{equation}

	The orthogonal part is discretized implicitly while the	non-orthogonal contribution is treated explicitly and added to the	matrix source term. After the resolution of the pressure Poisson equation, velocity field and conservative face flux $\phi$ are respectively updated using Equations \ref{eq:-20} and \ref{eq:-18} with the updated pressure field.

	\subsection{Ghost Fluid Method for pressure extrapolation}

	The GFM has been initially proposed by \citet{Fedwick1999} for sharp
	density handling in compressible flows. The method as been subjected
	to various extensions: \citet{kang2000}, \citet{Hong2007OnBC} or
	\citet{Lalanne2015}. The jump conditions \ref{eq:-8} and \ref{eq:-9}
	are used to derive interface-corrected interpolation schemes in the
	manner of \citet{Vukcevic2016}, where the procedure is given in details.
	First, mesh faces that share the interface are marked (Equation \ref{eq:-10})
	depending on the LS values $\psi_{P}$ and $\psi_{N}$ of an owner
	and its neighbor cells. Then, the coordinate vector of the interface
	$\boldsymbol{x}_{\Gamma}$ (Equation \ref{eq:-10-1}) is obtained by linear
	interpolation using distance weight $\lambda$ (Equation \ref{eq:-10-1-1})
	and owner and neighbor cell coordinates. Finally, the extrapolated
	pressure at the ghost cell is calculated using jump conditions \ref{eq:-8}	and \ref{eq:-9}. The ghost pressures are directly given in Equation \ref{eq:-11}
	and Equation \ref{eq:-12}, \textcolor{black}{\cite{Ferro2022}}.

	\begin{equation}
		\psi_{P}\psi_{N}<0\label{eq:-10}
	\end{equation}

	\begin{equation}
		\lambda=\frac{\psi_{N}}{\psi_{P}-\psi_{N}}\label{eq:-10-1-1}
	\end{equation}

	\begin{equation}
		\boldsymbol{x}_{\Gamma}=\boldsymbol{c}_{N}\text{+}\lambda\left(\boldsymbol{c}_{P}-\boldsymbol{c}_{N}\right)\label{eq:-10-1}
	\end{equation}

	\begin{equation}
		\begin{cases}
			p_{N}^{+}=\frac{\rho^{+}}{\rho^{*}}p_{d,N}+\left(1-\frac{\rho^{+}}{\rho^{*}}\right)p_{d,P}-\frac{\rho^{+}}{\rho^{*}}\mathcal{H} & P\;wet,\;N\;dry\\
			p_{P}^{-}=\frac{\rho^{-}}{\rho^{*}}p_{d,P}+\left(1-\frac{\rho^{-}}{\rho^{*}}\right)p_{d,N}-\frac{\rho^{-}}{\rho^{*}}\mathcal{H} & P\;wet,\;N\;dry
		\end{cases}\label{eq:-11}
	\end{equation}

\textcolor{black}{
	\begin{equation}
		\begin{cases}
			p_{N}^{-}=\frac{\rho^{-}}{\rho^{*}}p_{d,N}+\left(1-\frac{\rho^{-}}{\rho^{*}}\right)p_{d,P} + \frac{\rho^{-}}{\rho^{*}}\mathcal{H} & P\;dry,\;N\;wet\\
			p_{P}^{+}=\frac{\rho^{+}}{\rho^{*}}p_{d,P}+\left(1-\frac{\rho^{+}}{\rho^{*}}\right)p_{d,N}-\frac{\rho^{+}}{\rho^{*}}\mathcal{H} & P\;dry,\;N\;wet
		\end{cases}\label{eq:-12}
	\end{equation}
}

	Where $p_{d,N}$ and $p_{d,P}$ are the dynamic pressures
	at owner and neighbor cells of the interface face. $\rho^{*}=\rho^{-}\lambda+\rho^{+}\left(1-\lambda\right)$
	is the weighted average density as in \citet{Haghshenas2019}. $\rho^{-}$and
	$\rho^{+}$ are the density of light and heavy phases. Expressions
	\ref{eq:-11} and \ref{eq:-12} are used to modify the pressure gradient (Gauss
	and least square methods) and Laplacian operators for the pressure Poisson equation during the
	pressure-velocity coupling.

	\subsection{Relaxation Zone}

    An \textcolor{black}{explicit} relaxation zone method has been coded based on \citet{JacobsenRelax}. \textcolor{black}{In a relaxation zone, the solution is forced via an explicit correction. For free surface flows involved in marine applications, the target solution ($\psi$ and $u$) are usually calculated based on wave theory and depending on the desired sea state.} The method presented by \citet{JacobsenRelax} consists in applying the following \textcolor{black}{smoothing} equation to all cells within a specified zone for a given field $\xi$:
    	\begin{equation}
		\xi=(1-\omega)\xi_{cur}+\omega\xi_{tar}
	\end{equation}
	With $\xi_{tar}$ the desired value, $\xi_{cur}$ the current value and $\omega$ being a weight decaying exponentially within the zone and null outside of it 
	:
	\begin{equation}
		\omega=\frac{exp(\frac{d}{\lambda})-1}{e-1}
	\end{equation}

        \textcolor{black}{Where $d$ is the distance to the inlet of the zone and $\lambda$ the total length of the zone so that $w=0$ at the beginning of it and tends to $w=1$ at the end, imposing the desired solution when $d \rightarrow \lambda$. In this work, the relaxation zones are used for the last test case, in chapter \ref{KCS}, to damp induced waves so that numerical pollution or reflexion are avoided. The target solutions for $\psi$ and $\textbf{u}$ correspond to a flat sea state with a forced air/water current.}
        
	\subsection{Solver chart of \textit{LSFoam}}

\begin{figure}[h!]
		\centering
		\includegraphics[scale=0.45]{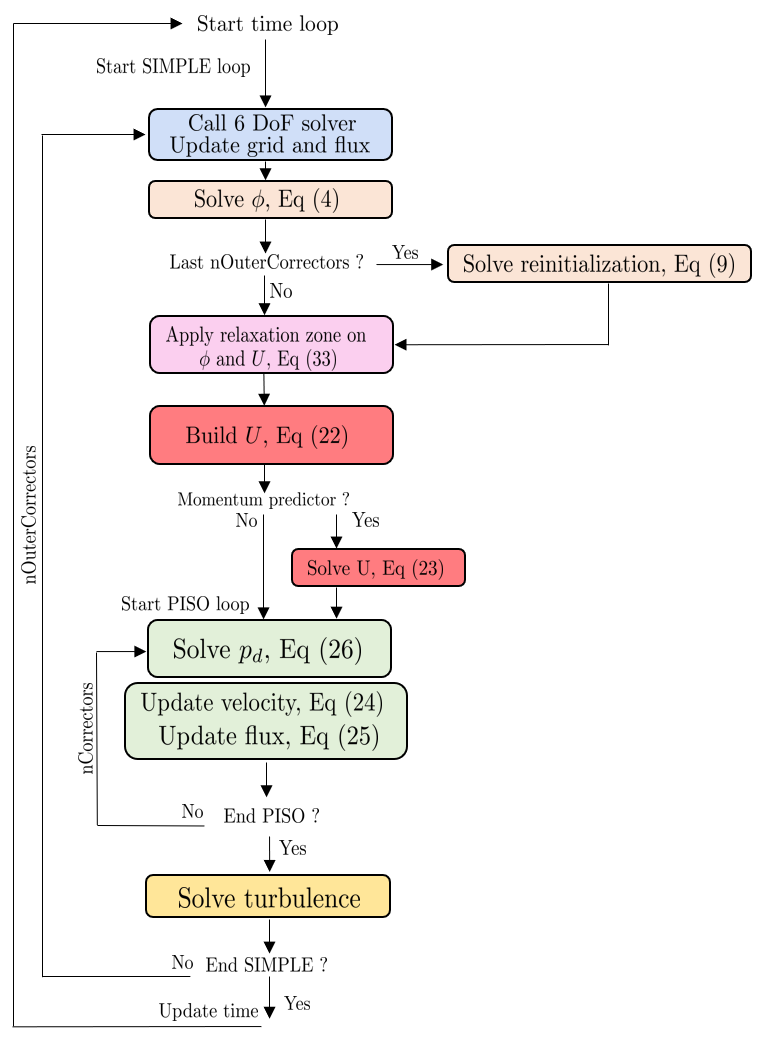}
		\caption{Segregated algorithm for \textit{LSFoam} solver. }
		\label{fig:Segregated-algorithm-for}
	\end{figure}

	The Level-Set, the momentum, and the pressure Poisson equations
	are solved in a segregated manner with the PIMPLE algorithm available
	in OpenFOAM. The PIMPLE algorithm is a combination of SIMPLE (\citet{Patankar1972})
	and PISO (\citet{Patankar1972,Issa1986}) algorithms. At the beginning
	of the time step, the SIMPLE loop starts. Grid and flux are updated knowing
	rigid body motions (if any) and Level-Set equation is solved. The reinitialization
	equation is solved only for the last SIMPLE corrector. \textcolor{black}{Explicit corrections 
        are operated for relaxation zones (if any)}. Then, interface
	faces are marked and fluid properties are updated. The pressure Poisson
	equation is solved iteratively within the PISO loop and, finally, the turbulence
	equations are solved. It has to be noticed that the $\textbf{H}$ operator is updated for each corrector of the PISO loop. The steps are resumed in Figure \ref{fig:Segregated-algorithm-for}.

         \clearpage

	\section{Test cases}
\begin{figure}[h!]
		\centering
		\includegraphics[scale=0.55]{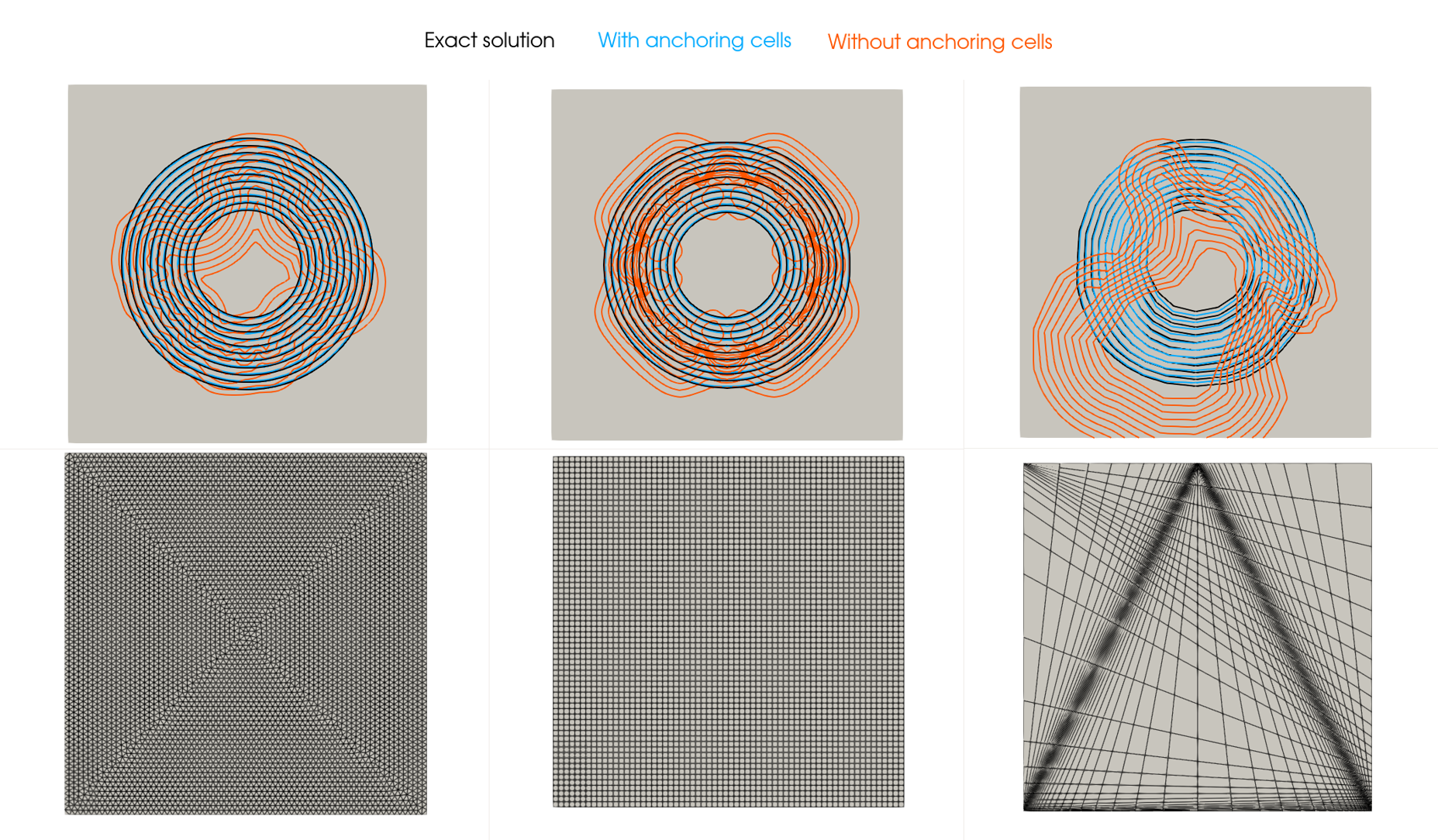}
		\caption{\centering Comparison of iso-contour of $\psi$ for the three grids at iteration $i=200$. Solution with anchoring cells in blue and without in orange.}
		\label{fig:psiResult}
	\end{figure}

        \textcolor{black}{\subsection{Assessment of enhanced reinitialization procedure}}\label{LS_testcase1}
        \textcolor{black}{        
         The performances of the enhanced reinitialization procedure, described in chapter \ref{LS}, is assessed using a dedicated numerical test case. Only the eikonal equation \ref{eq:FVMreini} is solved to recover the signed distance function, knowing a circular contour ($\psi^0=0$) of radius 0.25 m and with an arbitrary signed distance initialization  (-1 m inside the circle and 1 m outside it). The numerical domain is a 1m x 1m square. Three different 2D grids, shown in Figure \ref{fig:psiResult}, are generated: a 64x64 structured Cartesian grid, an unstructured grid with triangular cells and a structured distorted grid with a maximal non-orthogonality of 65° and a maximal aspect ratio of 70. The numerical error relative to the exact solution $\psi_{exact}$ is calculated using the $L_2$ norm, \cite{Kim2021} for each reinitialization iteration $i$: } 
         
        \textcolor{black}{   
	\begin{equation}
		L_2(i)= \left[ \overset{N_{cells}}{\underset{P}{\sum}} \left(\frac{(\psi_P^i-\psi_{P,exact})^2}{max(|\psi_{exact}|^2)N_{cells}}\right) \right]^{1/2}
	\end{equation}}

        \textcolor{black}{Where $N_{cells}$ is the total number of grid cells, $P$ the cell counter and i the iteration. }

        \textcolor{black}{\subsubsection{Use of anchoring cells}}
        
\begin{figure}[h!]
		\centering
		\includegraphics[scale=0.4]{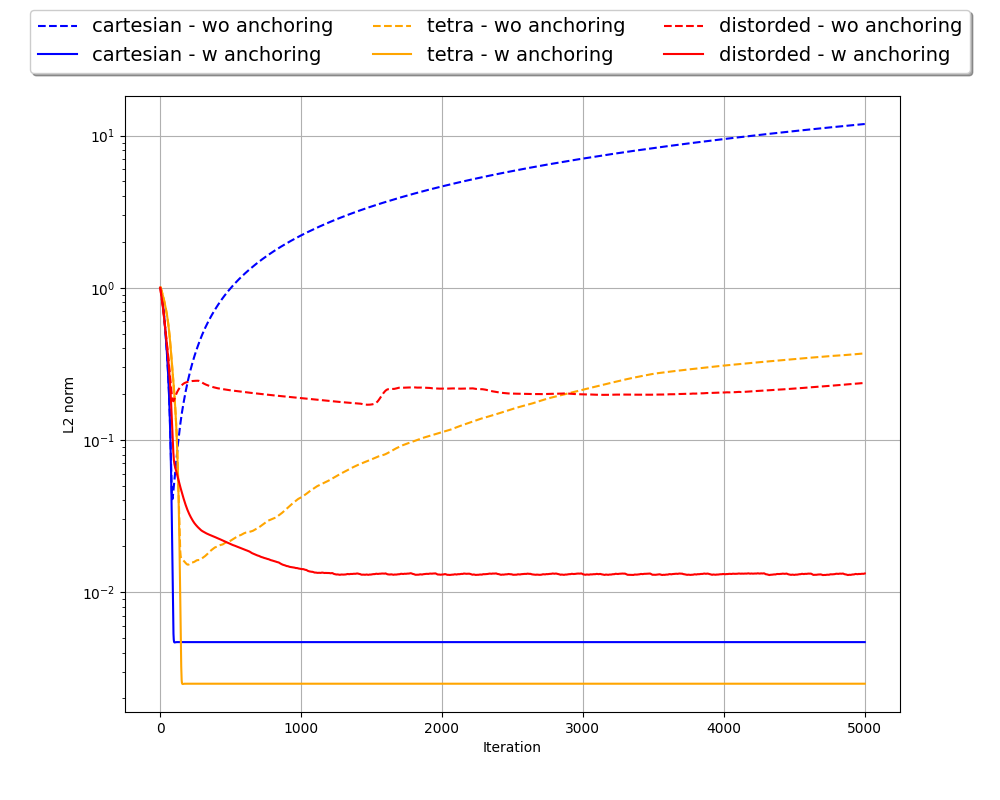}
		\caption{Evolution of $L_2$ norm during the reinitialization process for tree different 2D grids (Cartesian, tetrahedral and distorted) and w/wo anchoring cells. Dashed lines: without anchoring cells. Straight lines: with anchoring cells.}
		\label{fig:L2error}
    	\end{figure}

        \textcolor{black}{For each grids, the reinitialization equation \ref{eq:FVMreini} is solved for 5000 iterations with and without the anchoring cells strategy. Figure \ref{fig:L2error} shows the evolution of the $L_2(i)$ norm for the three grids, both with and without anchoring cells. The results indicate that without anchoring cells, the $L_2$ norm value starts to increase after a given amount of iterations, indicating an error accumulation due to interface displacements. The $L_2$ norm is one order of magnitude greater without the anchoring cells. In contrast, the method with anchoring cells exhibits monotonic convergence for the three grids. The convergence is slower for the distorted grid due to the nature of the grid itself. Figure \ref{fig:psiResult} shows the comparison of 11 $\psi$ iso-contours (from -0.1 m to 0.1 m) for both methods and for all the grids. The calculated iso-contours are plotted at $i=200$. The more iteration are solved, the more the Level-Set values without anchoring diverge. The exact desired analytical solution is also plotted in black. Without anchoring cells, the result exhibits}  \textcolor{black}{major difference compared to the analytical solution, whereas the use of anchoring cells provides a converged solution close to the analytical result.}
        
        \textcolor{black}{\subsubsection{Use of LTS approach}}

\begin{figure}[h!]
		\centering
		\includegraphics[scale=0.4]{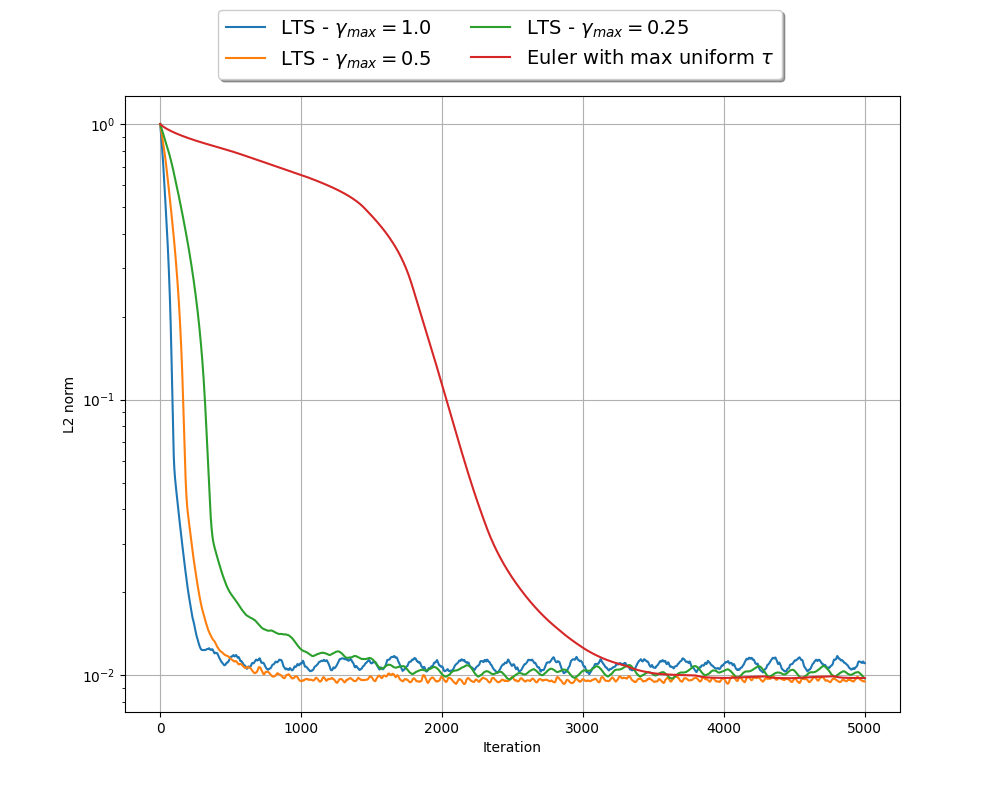}
		\caption{Evolution of $L_2$ norm during the reinitialization process for the distorted grid using a LTS approach (three Courant number $\gamma_{max}$ are compared) or with Euler scheme and the maximal admissible uniform spatial time step $\tau$.}
		\label{fig:L2errorSensi}
    	\end{figure}

        \textcolor{black}
        {
        Another improvement of the original method of \cite{Sussman1998} proposed in this work is the use of LTS approach for the discretization of the temporal term in Equation \ref{eq:FVMreini}, replacing the Euler scheme with a uniform spatial time step $\tau$. For the original approach and with industrial meshes containing a wide range of different sizes, the eikonal equation convergence speed would be determined by the CFL condition for the smallest cell. In contrast, a LTS approach allows to increase the convergence speed by maximizing locally the time step (Equation \ref{eq:tauVariable}) based on the user defined reinitialization Courant number $\gamma_{max}$. Figure \ref{fig:L2errorSensi} shows the evolution of the $L_2$ norm during the reinitialization (5000 iterations) for the distorted grid using both methods. For the original one, the spatial time step $\tau$ is maximized while avoiding any divergence of the resolution process. In both cases, the eikonal equation is solved implicitly as described in chapter \ref{LS}. The results show that the LTS approach allows to drastically increase the convergence speed (by approximately a factor 10, when $\gamma_{max} = 1.0$) while maintaining the accuracy and the stability, demonstrating the efficiency of the present approach for complex meshes.}

         \clearpage
 
         \subsection{The rising bubble test case, \citet{Hysing2007}}

	The rising bubble test case is a 2D benchmark from \citet{Hysing2007}.
	This test case has been simulated by many authors: \citet{Zuzio2011},
	\citet{Klostermann2013}, \citet{Balcazar2016}, \citet{Patel2017},
	\citet{Gamet2020}. It consists of simulating a single rising bubble in a quiescent
	liquid. The domain sizes in the x and y directions are 1 m and 2 m. A
	bubble with a diameter of $D=0.5$ m is initially centered at coordinates
	(x,y)=(0.0, 0.5). Top and bottom boundaries are non-slipping walls,
	whereas lateral walls are slipping ones. Two configurations, TC1 and
	TC2, are simulated and resumed in table \ref{tab:TC1-and-TC2:}. The
	Reynolds number $R_{e}$, the Etvs number $E_{O}$ and the capillary
	number $C_{a}$ are defined as:

	\begin{equation}
		\begin{array}{ccc}
			R_{e}=\frac{\rho_{1}U_{g}D}{\mu_{1}} & \;\;\;\;E_{O}=\frac{\rho_{1}U_{g}^{2}D}{\sigma} & \;\;\;\;C_{a}=\frac{E_{O}}{R_{e}}\end{array}\label{eq:-13}
	\end{equation}

	Where $U_{g}=\sqrt{gD}$. Four levels of Cartesian grid
	refinements are studied: 32x64, 64x128, 128x256 and 256x512. Simulations
	are performed using an adaptive time step based on a maximal Courant
	number value of 0.05. The solver is set in PIMPLE (\citet{Issa1986})
	mode, and the number of loops depends on the residual of each iteration. The PIMPLE algorithm stops when the calculated residuals are lower than $\epsilon_{p_{d}} < 10^{-5}$, $\epsilon_\textbf{U} < 10^{-4}$ and $\epsilon_\psi < 10^{-6}$. The convection term in the momentum equation is
	discretized with a second order upwind scheme (\textit{linearUpwind}) and the dynamic pressure gradient
	is calculated using a least square method corrected at the interface (Ghost Fluid Method). Time advancement is achieved with the 2nd order backward scheme (\textit{backward}).
	The reinitialization equation is solved 10 times at each time step with a reinitialization
	Courant number of 0.75. This procedure allows to fully recover the signed distance property of
	the Level-Set function. The characteristic length $\epsilon$ is chosen
	as $2 \Delta z(\boldsymbol{x})$. The mass error is less
	than $10^{-3}$ \%.

	\begin{table*}[h!]
		\centering
		\begin{tabular}{|c|c|c|c|c|c|c|c|c|c|c|c|}
			\hline
			case & $\rho_{1}$ & $\rho_{2}$ & $\mu_{1}$ & $\mu_{2}$ & $g$ & $\sigma$ & $R_{e}$ & $E_{O}$ & $C_{a}$ & $\frac{\rho_{1}}{\rho_{2}}$ & $\frac{\mu_{1}}{\mu_{2}}$\tabularnewline
			\hline
			\hline
			TC1 & 1000 & 100 & 10 & 1 & 0.98 & 24.5 & 35 & 10 & 0.286 & 10 & 10\tabularnewline
			\hline
			TC2 & 1000 & 1 & 10 & 0.1 & 0.98 & 1.96 & 35 & 125 & 3.571 & 1000 & 100\tabularnewline
			\hline
		\end{tabular}
		\caption{TC1 and TC2: Physical properties and similarity parameters}
		\label{tab:TC1-and-TC2:}
	\end{table*}

	Two quantities are used to compare the simulation results
	with the reference data of \citet{Hysing2007}: the position of the
	bubble center of mass (Equation \ref{eq:-14}) and the bubble vertical velocity
	(Equation \ref{eq:-14-1}).

	\begin{equation}
		\boldsymbol{X}_{c}=\frac{\int\alpha\boldsymbol{x}dx}{\int\alpha dx}\label{eq:-14}
	\end{equation}

	\begin{equation}
		V_{y}=\frac{\int\alpha v_{y}dx}{\int\alpha dx}\label{eq:-14-1}
	\end{equation}

\begin{figure}[h!]
		\centering
			\includegraphics[scale=0.49]{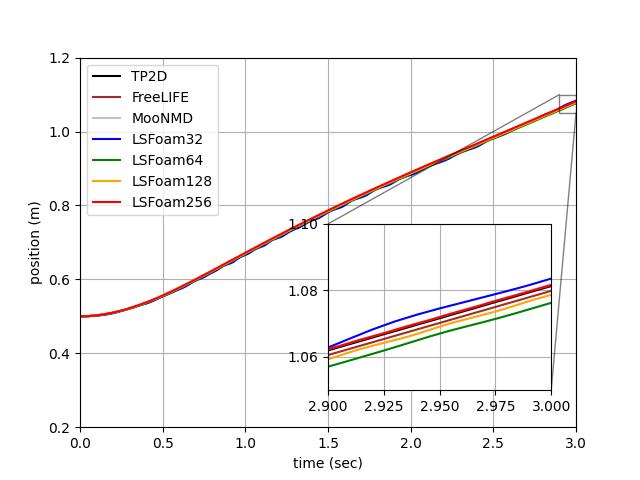}
			\includegraphics[scale=0.49]{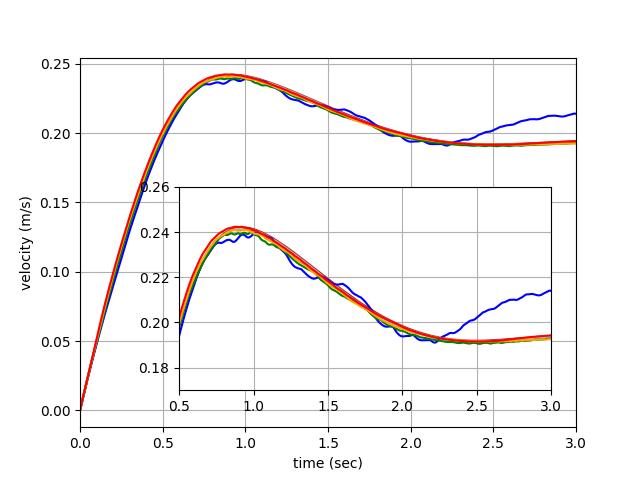}
		\caption{\centering Case TC1 - Bubble center of mass coordinates and rising velocity comparison with data of \citet{Hysing2007} \label{fig:Case-TC1-:}.}
	\end{figure}

\begin{figure}[h!]
		\centering
			\includegraphics[scale=0.49]{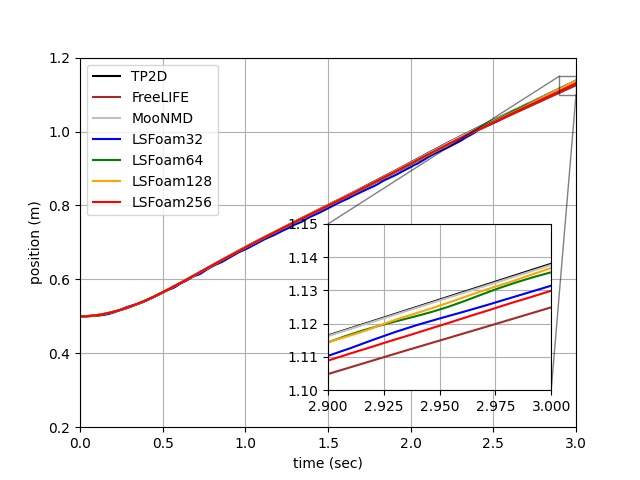}
			\includegraphics[scale=0.49]{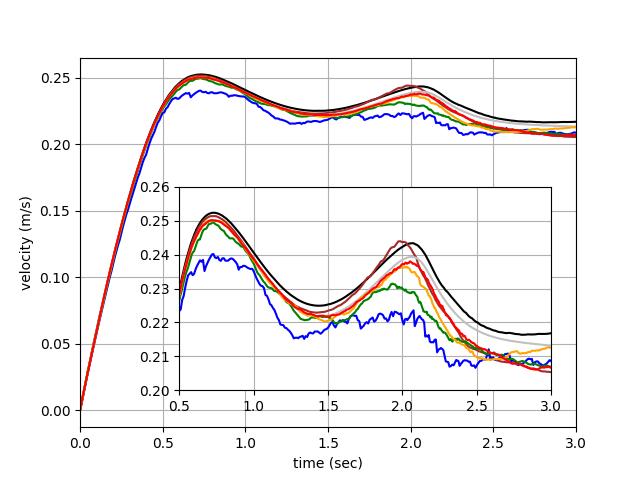}
			\caption{\centering Case TC2 - Bubble center of mass coordinates and rising velocity comparison	with data of \citet{Hysing2007} \label{fig:Case-TC2-:}.}
	\end{figure}

\begin{figure}[h!]
		\centering
			\includegraphics[scale=0.49]{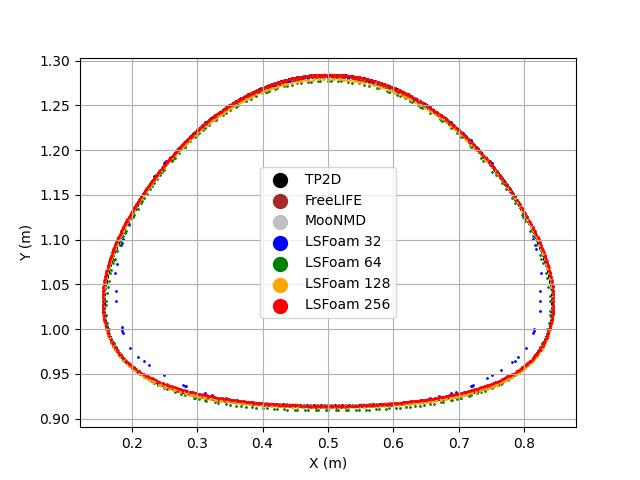}
			\includegraphics[scale=0.49]{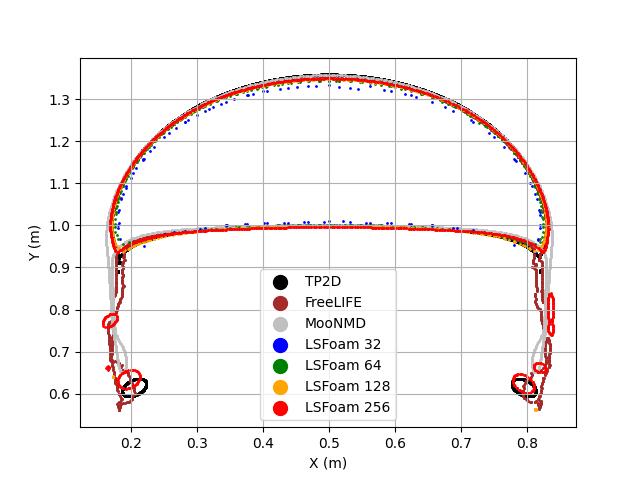}
			\caption{\centering Case TC1 and TC2 - Bubble shape comparison with data of \citet{Hysing2007}.
			\label{fig:bubble shape-:}}
	\end{figure}

    Where $\alpha$ is the volume fraction as defined in equation
	\ref{eq:-8}. The bubble center of mass evolutions are presented in
	Figure \ref{fig:Case-TC1-:} for each level of grid refinement. Overall
	results are in relatively good agreement \textcolor{black}{with the reference codes named MoonNMD, FreeLIFE and TP2D from} \citet{Hysing2007},
	especially for the two finest grids. The rising velocity results are
	presented in Figure \ref{fig:Case-TC2-:}. As for the bubble center
	of mass, there is a good agreement between the presented results and the data
	of \citet{Hysing2007}. The results tend to be closer to the reference
	data with grid refinements. The numerical results can also be compared
	qualitatively by examining the bubble shape for the two test cases, as
	presented in Figure \ref{fig:bubble shape-:}.

    \newpage
\subsection{Rayleigh-Taylor instability, \citet{Puckett1997}}
\begin{figure}[h!]
    		\centering
    		\includegraphics[scale=0.35]{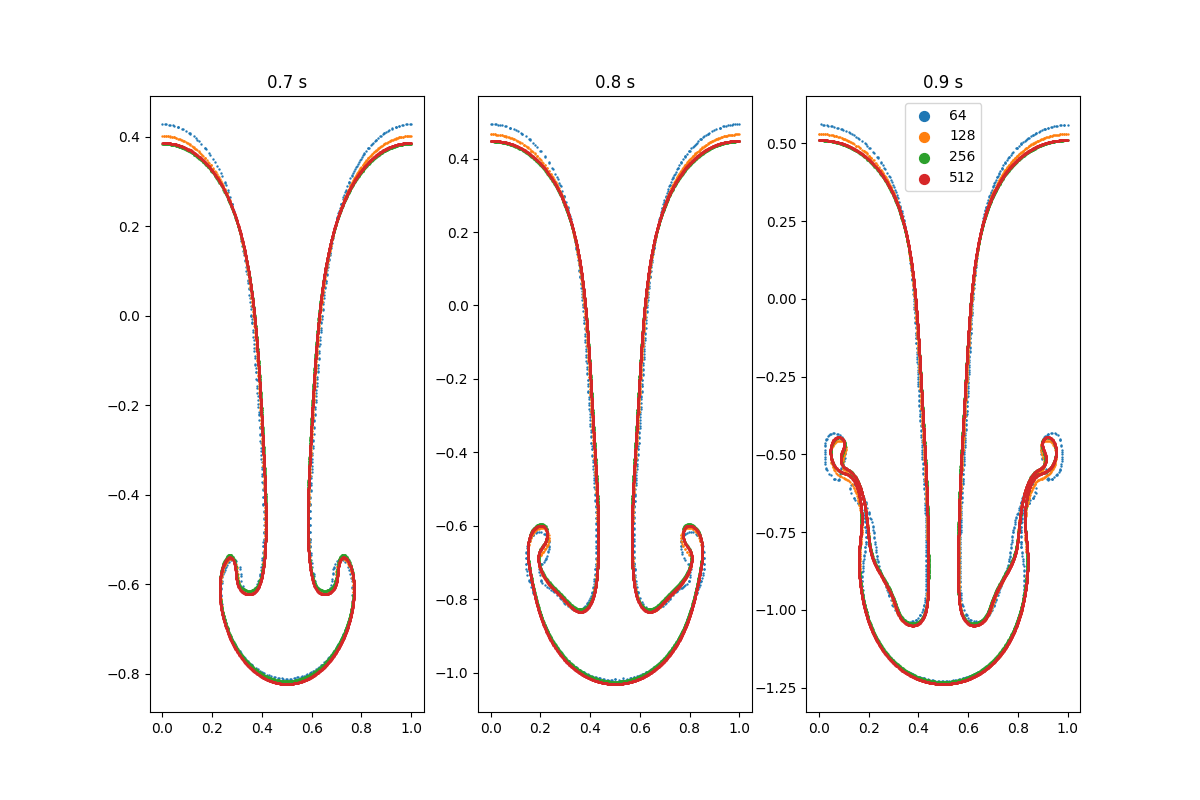}
    		\caption{Rayleigh-Taylor instability. Shape comparison for mesh sensitivity results.}
    		\label{Ray and Tay1}
    \end{figure}

\begin{figure}[h!]
		\centering
	    \includegraphics[scale=0.35]{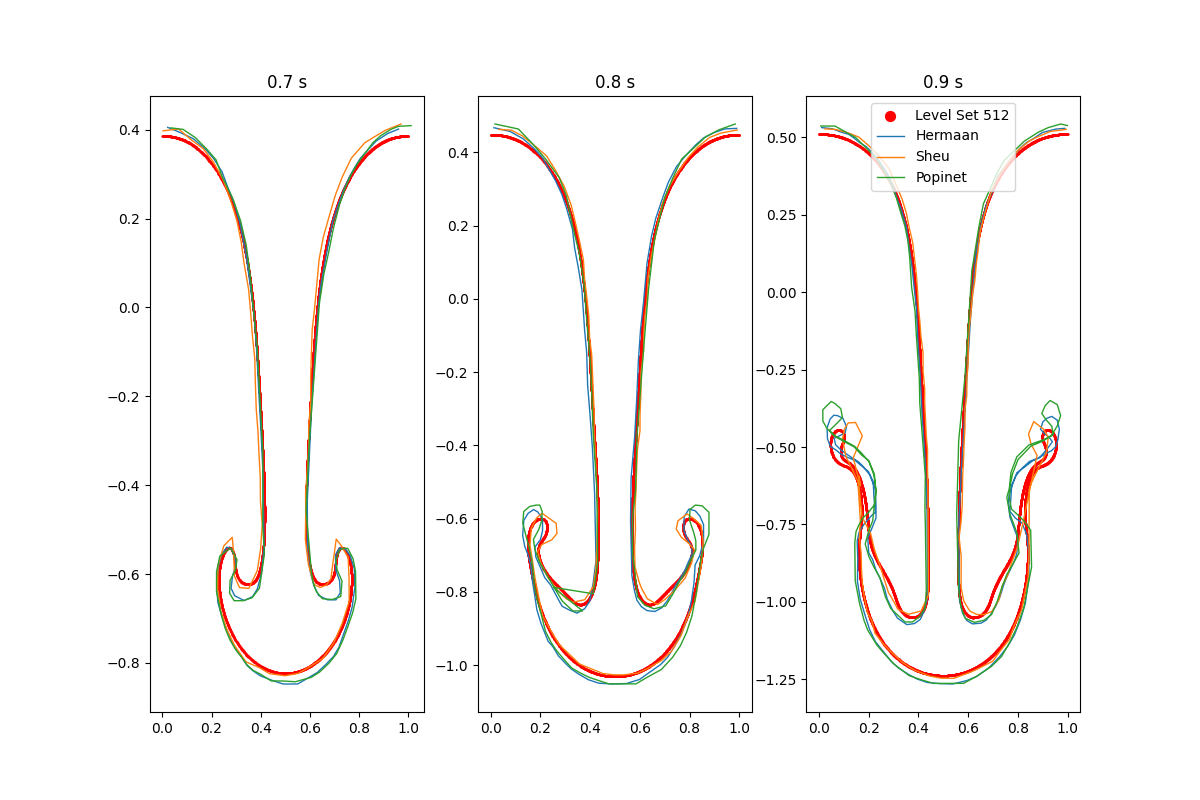}
		\caption{Rayleigh-Taylor instability. Shape comparison between finest grids and reference solutions.}
		\label{Ray and Tay2}
	\end{figure}

	The Rayleigh-Taylor instability test case is a popular 2D numerical benchmark for complex multiphase flows. The problem has been initially proposed by \citet{Puckett1997} and simulated by many authors during the last two decades: \cite{popinet:hal-01445441}, \citet{Herrmann2008}, \citet{Sheu2009}, \citet{Talat2018} or \citet{Kim2021}. The case consists in modeling two fluids with different density, $\rho_{1}=1.225$ and $\rho_{2}=0.1694 kg/m^3$, but with the same viscosity $\mu=0.00313 Pa.s$ in a rectangular domain measuring 1 m x 4 m. The heavier fluid is positioned above the lighter one, and the interface is initially shaped with a sinusoidal perturbation having an amplitude of 0.05 m. The top and bottom boundaries are non-slipping walls, whereas the lateral ones are slipping walls. Four levels of Cartesian grid refinements are studied: 64x256, 128x512, 256x1024 and 512x2048. Simulations are performed using an adaptive time step based on a maximal Courant number value of 0.2. The numerical settings are identical to the rising bubble test cases. Results are presented in Figure \ref{Ray and Tay1} and Figure \ref{Ray and Tay2} for mesh sensitivity and comparison with reference data. The mushroom shape tends to converge with grid refinement. The comparison with reference results shows a medium level of agreement, except with the data of \citet{Sheu2009} where the results are closer. The mass variation relative error is below $10^{-4}$.

\newpage

\subsection{Flow around an Ogee crest, \cite{erpicum2018experimental}}
	
The flow around a scale model of an Ogee spillway crest is reproduced. The W2 Ogee spillway geometry is modeled with a design head $H_d$ of 10 cm (corresponding to case W2 of \cite{PELTIER2018128}, \cite{erpicum2018experimental}) and have been built with the following equations from \cite{OGEE_geometry_eq} and \cite{OGEE_CFD}:

\begin{equation}
\left(\frac{x}{H_d} + 0.2418 \right)^2 + 	\left(\frac{z}{H_d} +0.1360 \right)^2 = 0.04^2 \; \text{for} \; 
	- 0.2818 \le \frac{x}{H_d} \le -0.276 \
\end{equation}
	\begin{equation}
	\left(\frac{x}{H_d} + 0.105 \right)^2 + 	\left(\frac{z}{H_d} +0.219\right)^2 = 0.2^2 \;  \text{for} \; 
	- 0.2276 \le \frac{x}{H_d} \le -0.175 
\end{equation}	
\begin{equation}
 \left(\frac{x}{H_d}\right)^2 + 	\left(\frac{z}{H_d} +0.5\right)^2 = 0.5^2 
 \;  \text{for} \;  
 -0.175  \le \frac{x}{H_d} \le 0 
\end{equation}	
\begin{equation}
 \frac{z}{H_d} = -0.5 * \left(\frac{x}{H_d}\right)^{1.85}  
 \; \text{for} \; \frac{x}{H_d} \ge 0
\end{equation}	

\begin{figure}[h!]
		\includegraphics[scale=0.4]{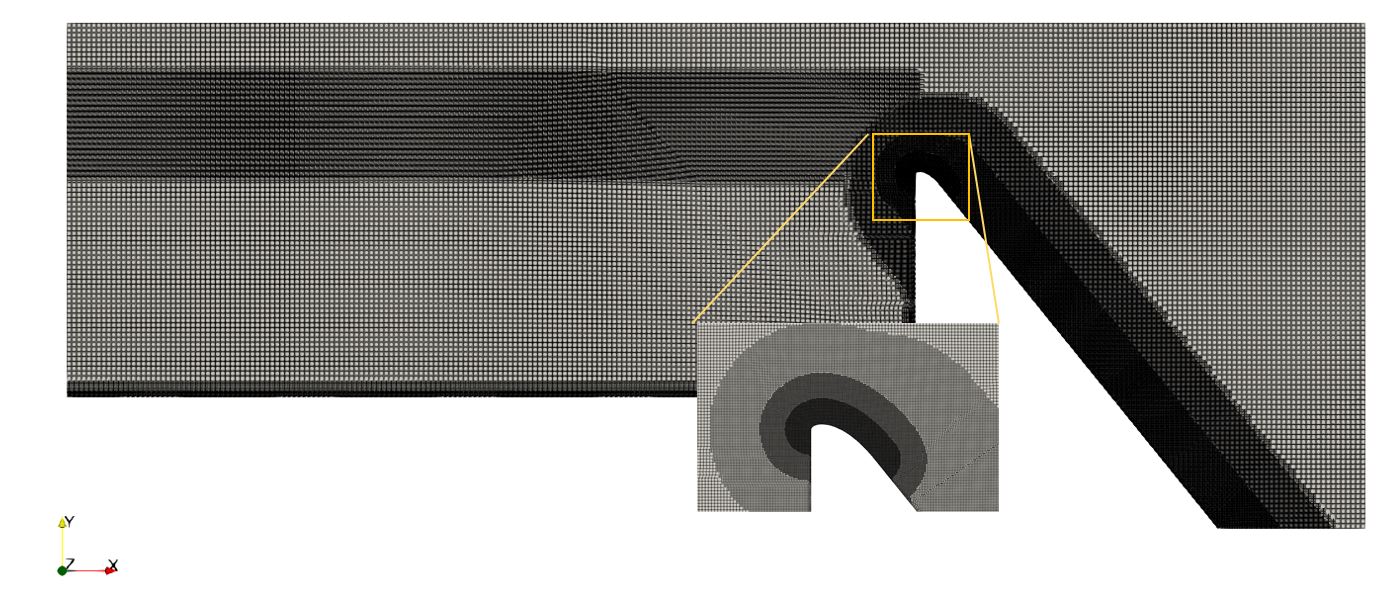}
		\centering
		\caption{Ogee spillway simulation - Illustration of the two-dimensional mesh with 200k cells.}
		\label{Ogee mesh}
	\end{figure}

\begin{figure}[h!]
    	\centering
		\includegraphics[scale=0.7]{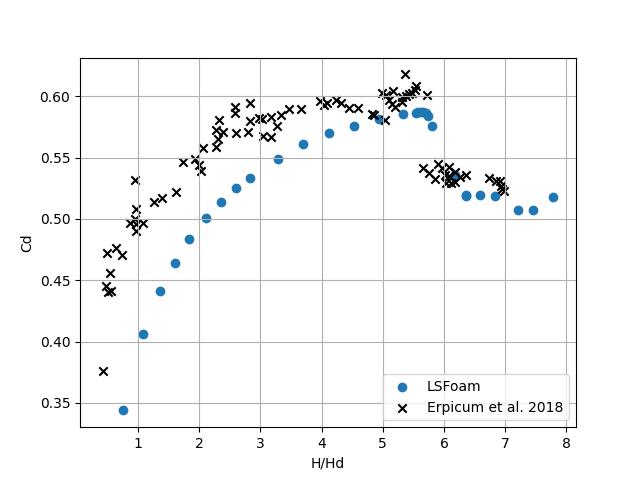}
		\caption{Discharge coefficients for different Head ratio with the W2 geometry.}
		\label{OGEE CD}
	\end{figure}

With x = 0 corresponding to the highest point of the Ogee spillway. The 2D numerical domain, illustrated in Figure \ref{Ogee mesh}, is composed of an upstream tank, the Ogee spillway, and a discharge zone. At the inlet, a Robin boundary condition is implemented, imposing the velocity of the water phase while applying a zero-gradient condition for the air phase. The reference pressure is imposed at the top through an atmospheric boundary condition. Zero-gradient conditions are applied to the right patch while the remaining patches are defined as wall types. The mesh sensitivity study didn't show any significant variation in the results. Therefore, the results are presented for a single unstructured grid generated with \textit{snappyHexMesh} and composed of 200k cells. \textcolor{black}{The maximum mesh non-orthogonality is 69° and the maximum aspect ratio equals 729}. The turbulence is solved with the EARSM turbulence model of \citet{Hellsten}. Regarding the numerical settings, the Euler scheme is used for time derivatives since only the steady-state is of interest. The convection terms are discretized with the 2nd order upwind scheme (\textit{linearUpwind}) with a limited gradient (\textit{cellLimited Gauss linear 1}). The \textit{MUSCL} scheme (\citet{vanLeer1979}) is used to discretize the Level-Set convective term. The gradients are discretized with the Gauss linear scheme, except for the pressure one, which is discretized with the least square scheme. The time step is fixed at 10 ms, leading to maximal Courant numbers around 100, and the PIMPLE algorithm stops when the calculated residuals are lower than $\epsilon_{p_{d}} < 10^{-5}$. For such a flow, the head $H$ is defined by the water depth $h$ relative to the crest corrected y a kinetic energy term:

\begin{equation}
	H = h + \frac{Q^2}{2gB^2(h + h_{uf})}
\end{equation}	
Where $Q$ is the discharge ($m^3/s$), $B$ the spillway width, and $h_{uf}$ the height of upstream face of the spillway. A sensitivity analysis on the water velocity at the inlet patch is performed for determining the discharge coefficient $C_D = \frac{Q}{B\sqrt{2gH^3}}$ relative to the head ratio $H/H_d$. The results are compared to the experimental data of \cite{erpicum2018experimental}. The water level is initialized at the crest, and the water velocity is ramped during 5 s. The simulations are conducted for a sufficient time (50 s) to ensure a stabilized state. The discharge coefficient relative to head ratio is shown in Figure \ref{OGEE CD}. The overall trend is in medium agreement with experimental data, and the discharge coefficient is underestimated. However, the flow separation near $H/H_d = 5.5$ is well captured. The deviations from the experimental data are likely caused by turbulence and/or 3D effects. More detailed investigations are out of the scope of this study. It has to be noticed that no drift of the water level has been observed, showing that the present method, at least for this test case, is mass conservative.

	\clearpage
	
	\subsection{Tridimensional dambreak simulation with a square cylinder obstacle,
		\citet{GomezGesteira2013}}

\begin{figure}[h!]
		\includegraphics[scale=0.2]{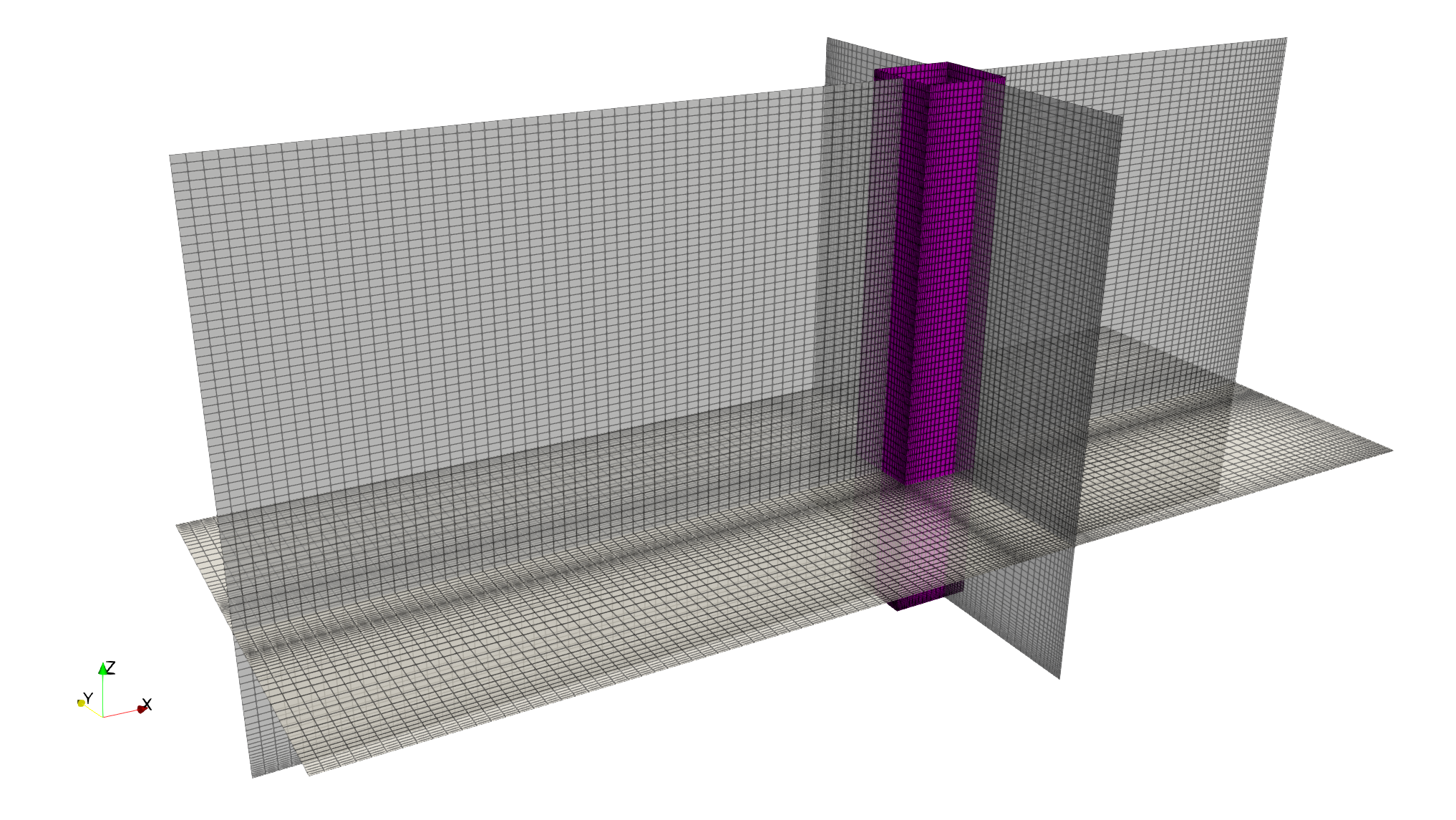}
		\centering
		\caption{\centering View of the finest non-uniform Cartesian grid built with \textit{blockMesh} - 3 M cells. The square obstacle is colored in purple.}
		\label{Dambreak mesh}
	\end{figure}

\begin{figure}[h!]
		\centering
		\includegraphics[scale=0.25]{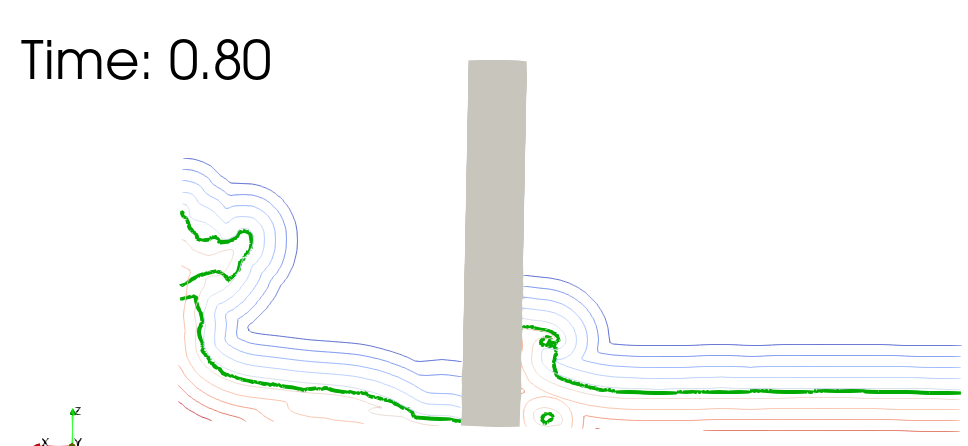}
		\includegraphics[scale=0.25]{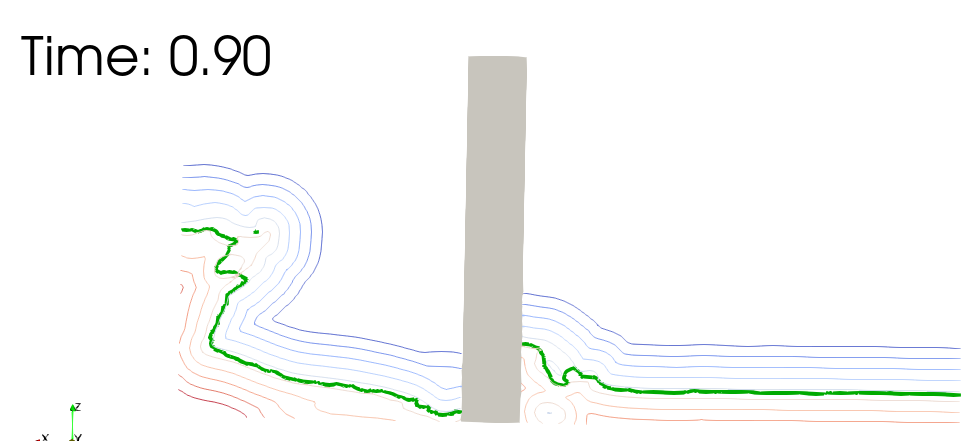}
		\includegraphics[scale=0.25]{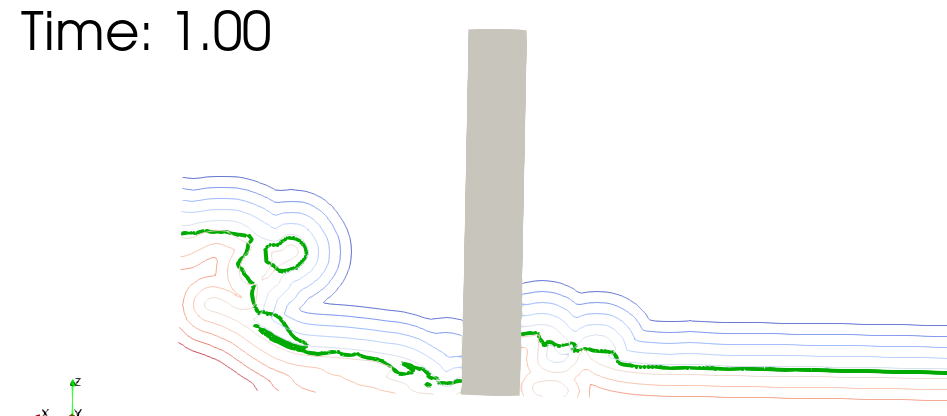}
		\includegraphics[scale=0.25]{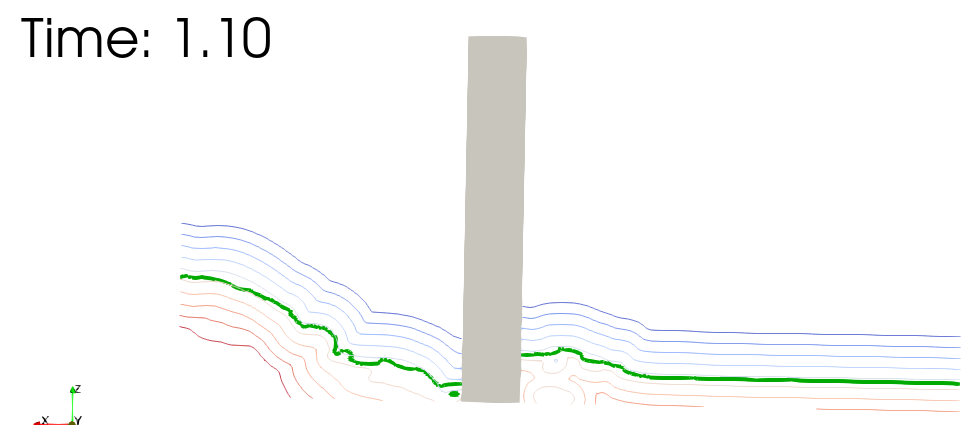}
		\includegraphics[scale=0.5]{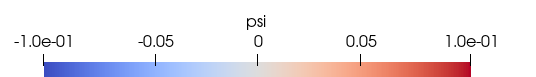}
		\caption{Level-Set contours - from -0.1 to 0.1 - $\psi(\boldsymbol{x})=0$ in green. \label{LS contours Dam}}
	\end{figure}

\begin{figure}[h!]
		\centering
		\includegraphics[scale=0.4]{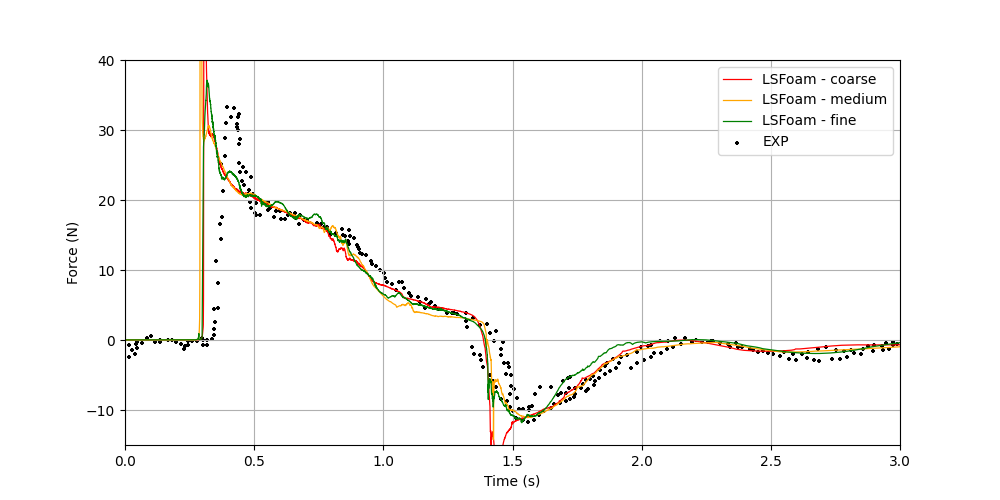}
		\includegraphics[scale=0.4]{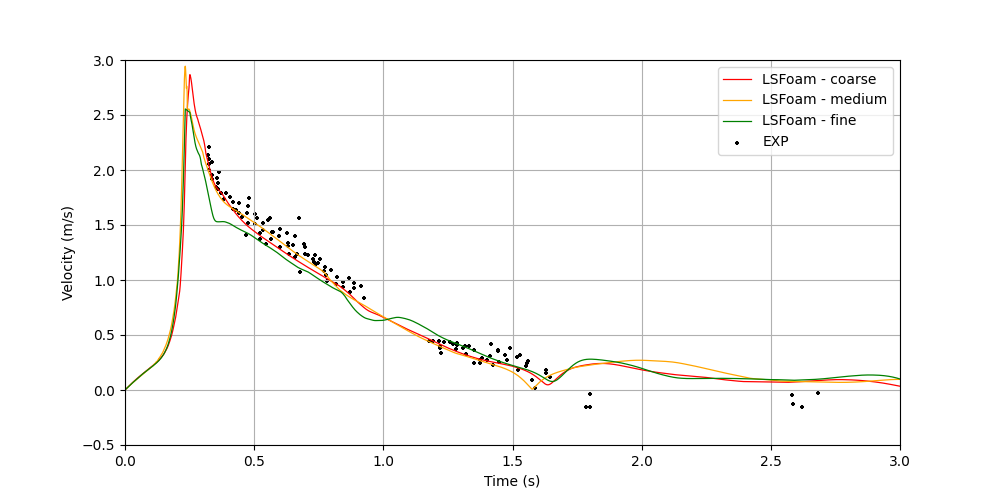}
		\caption{Force and velocities results for the 3 grids, and comparison with the experimental data \citet{GomezGesteira2013}. \label{Dam Break Res}}
	\end{figure}

\begin{figure}[h!]
		\centering
		\includegraphics[scale=0.22]{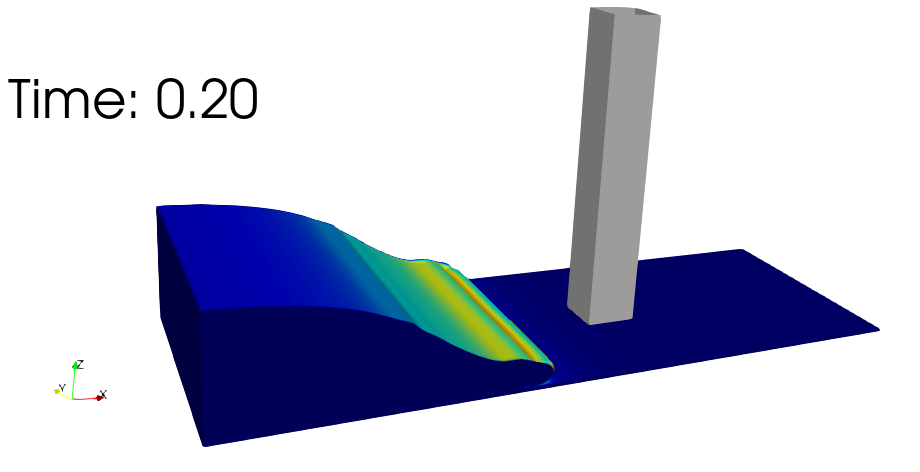}
.		\includegraphics[scale=0.22]{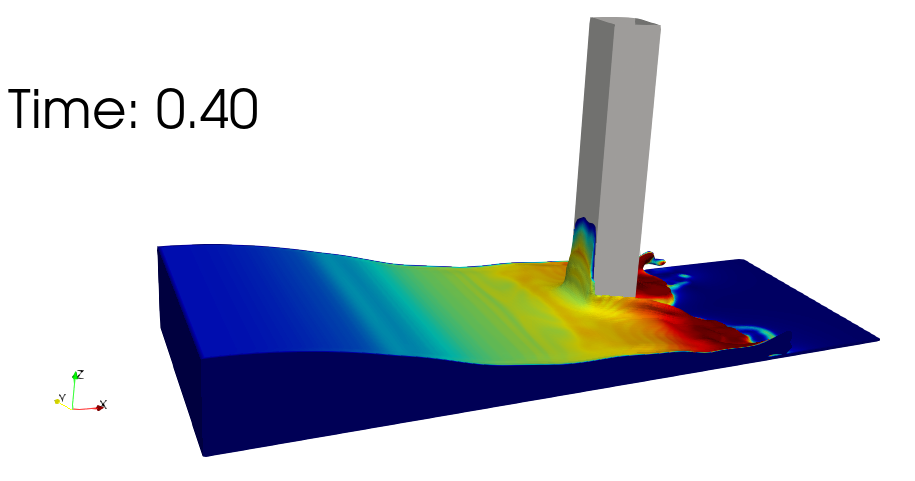}
		\includegraphics[scale=0.22]{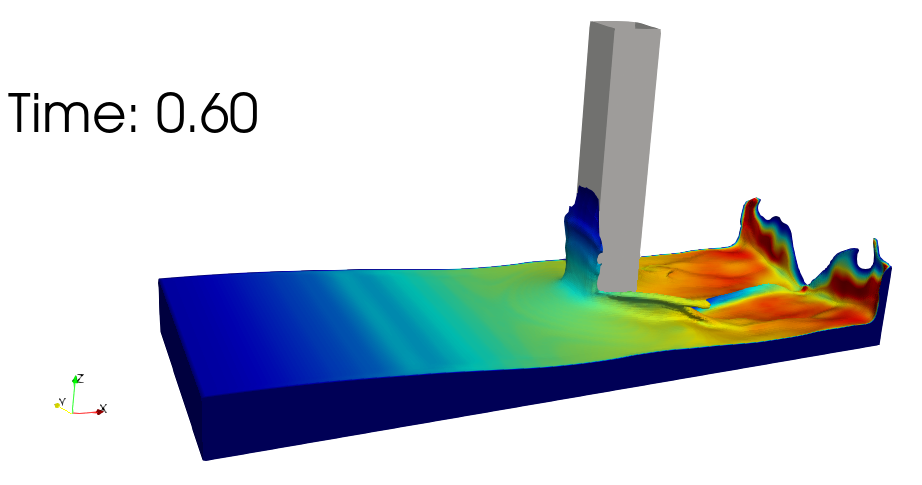}
		\includegraphics[scale=0.22]{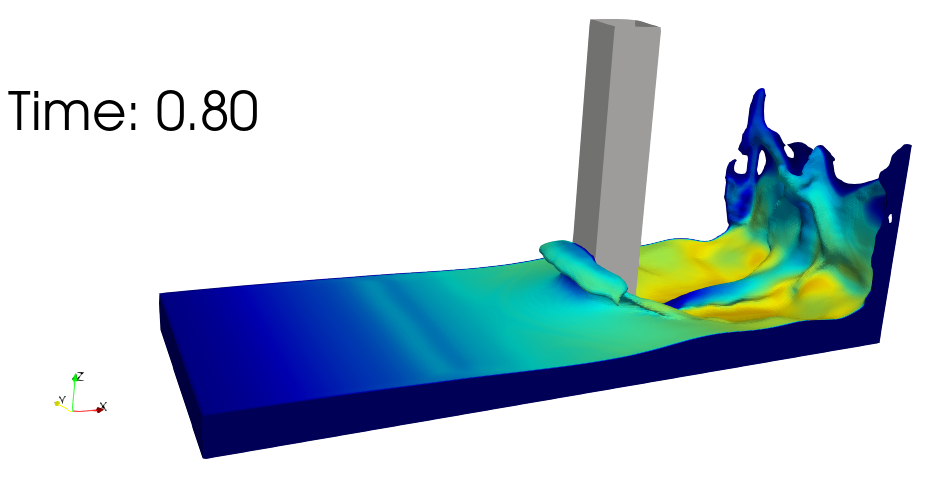}
		\includegraphics[scale=0.22]{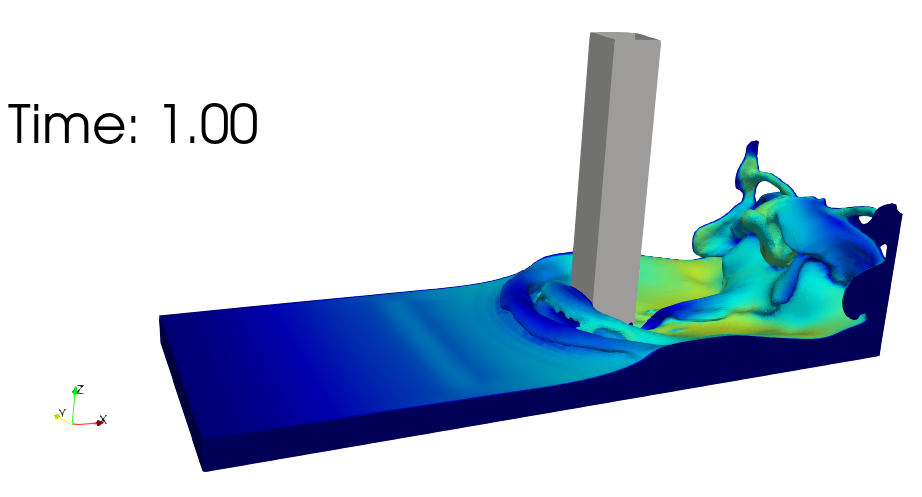}
		\includegraphics[scale=0.22]{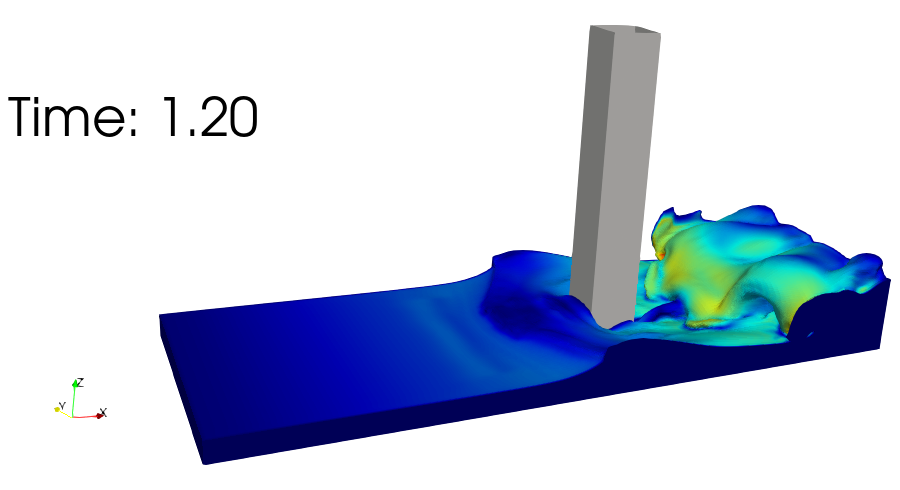}
		\includegraphics[scale=0.22]{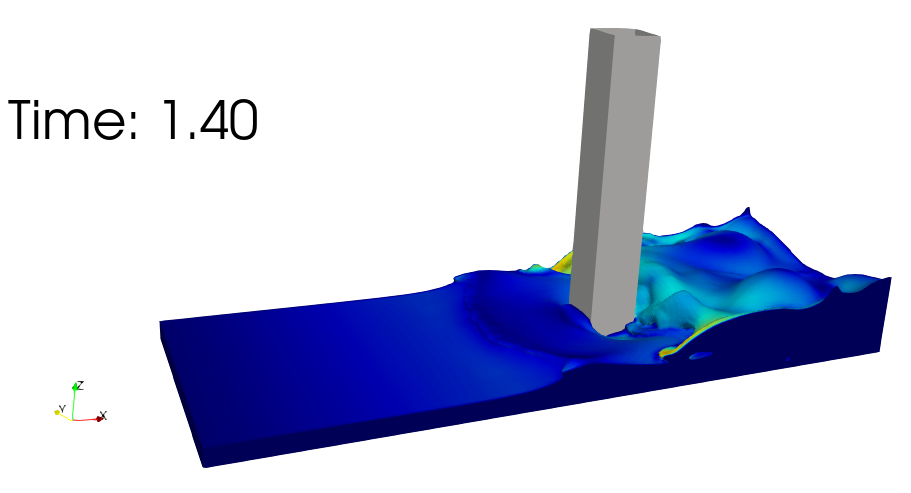}
		\includegraphics[scale=0.22]{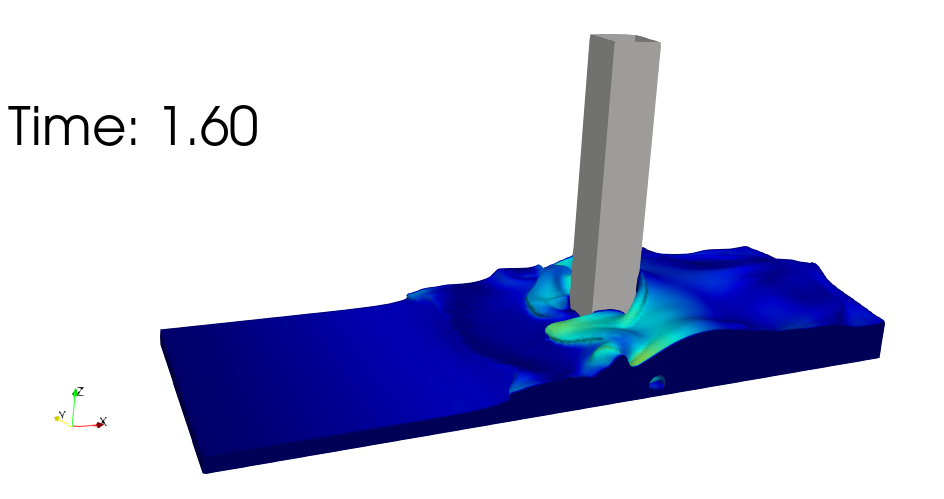}
		\includegraphics[scale=0.4]{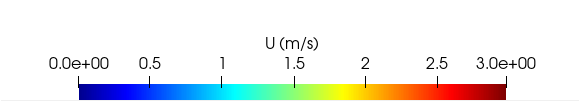}
		\caption{Free surface motions. \label{Free Surface Motion}}
	\end{figure}

	This test case consists of simulating the fall of a water column and
	the impact of the generated wave on a square cylinder obstacle.
	Numerical facilities are described by \citet{GomezGesteira2013} and
	\cite{Ferro2022}. A water depth of 7.5 mm is initially placed beyond the gate (based on \cite{Vukcevic2014}). Force and fluid velocity
	in front of the obstacle measurements have been performed and are
	used for comparison with numerical results. The velocity probing point
	is located 146 mm in front of the obstacle, in the mid-plane, and 26
	mm above the floor. Three mesh grids have been generated with \textit{blockMesh}
	and are composed of: 130 k (coarse), 300 k (medium), and 3 M (fine)
	hexahedral cells. The finest one is illustrated Figure \ref{Dambreak mesh}. Boundary conditions are wall types except for the
	top boundary, where an atmospheric boundary condition is simulated (\textit{totalPressure}	and \textit{pressureInletOutletVelocity}). Simulations are carried out using
	an adaptive time step based on a maximal Courant number value of 0.25
	for the interface and 0.5 for the remaining domain. Turbulence is solved with the
	EARSM turbulence model of \citet{Hellsten}. Convection
	terms are discretized with the 2nd order upwind scheme (\textit{linearUpwind})
	with a limited gradient (\textit{cellLimited Gauss linear 1}). Gradients are
	discretized with the Gauss linear scheme, except the pressure one, which is discretize	with the least square scheme. The \textit{MUSCL} schemes (\citet{vanLeer1979})
	is used to discretize the Level-Set convective term. Finally, the 2nd
	order backward scheme is chosen for time derivatives, and the PIMPLE algorithm stops when the calculated pressure residual is lower than $\epsilon_{p_{d}} < 10^{-5}$. Force and velocity
	histories are presented in Figure \ref{Dam Break Res}. The
	offset in the first peak is also observed by \cite{Vukcevic2014} and \cite{Ye2O20}. It may be caused by the fact that, experimentally, the gate can’t be instantly removed. Overall, the results are relatively similar and are in agreement with the experimental data, even though the
	peaks are slightly overestimated. 
	 Regarding velocity measurements in front
	of the obstacle, the trend is properly captured. Free surface motions
	are detailed in Figure \ref{Free Surface Motion}. Figure \ref{LS contours Dam}
	represents the Level-Set contours for the finest grid and illustrates that
	the Level-Set distance property is well preserved. In this case, where the free surface motions are complex, the relative error of mass variation is inferior to 1\%.

	\clearpage

	\subsection{Gothenburg workshop, 2010, KVLCC2 steady resistance, \citet{Larsson2014}} \label{KCS}

\begin{figure}[h!]
			\centering
			\includegraphics[scale=0.15]{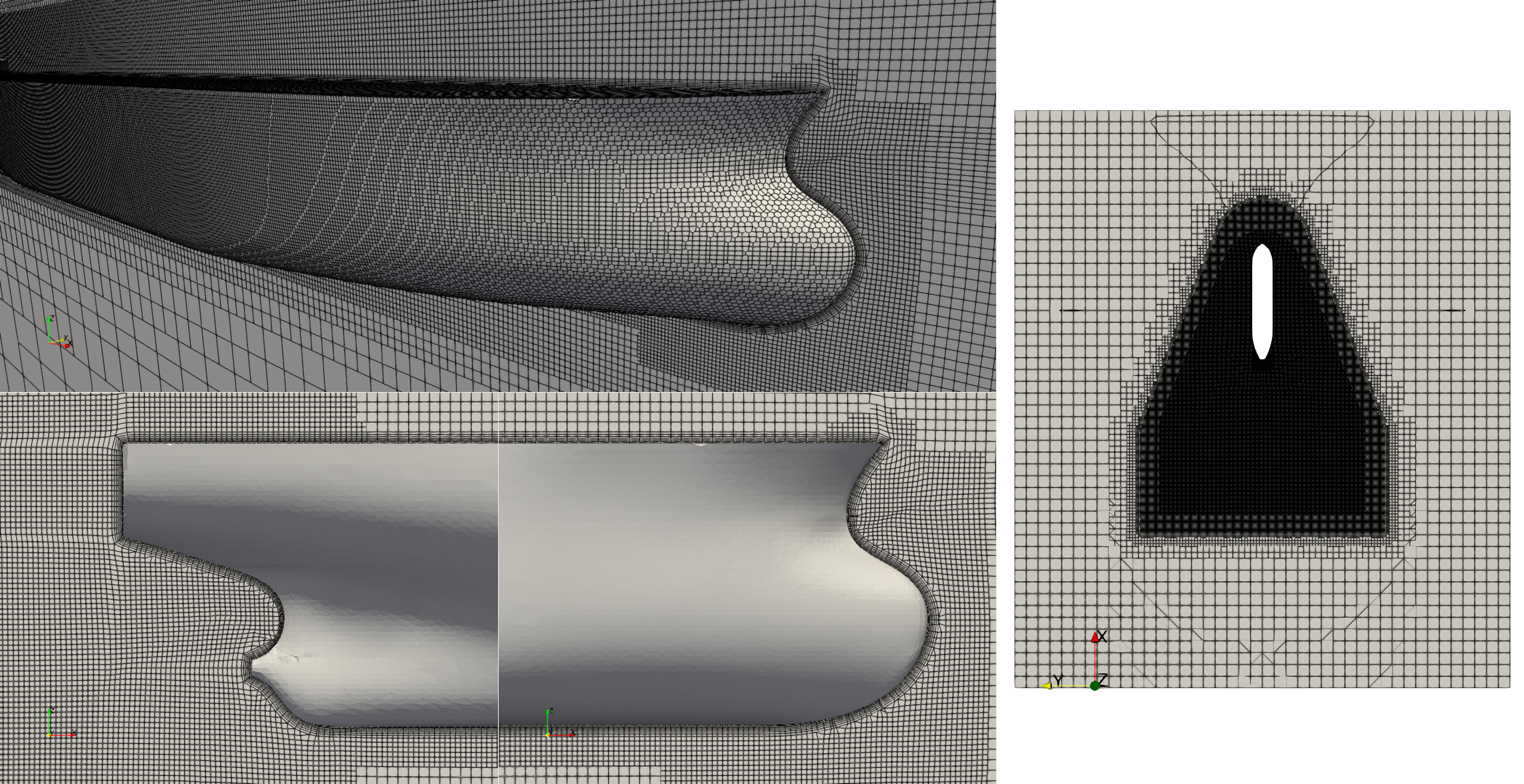}
			\caption{KVLCC2 mesh obtained with snappyHexMesh - very fine grid. 5.5 M cells.\label{fig:KVLCCmesh}}
		\end{figure}

\begin{figure}[h!]
			\centering
			\includegraphics[scale=0.5]{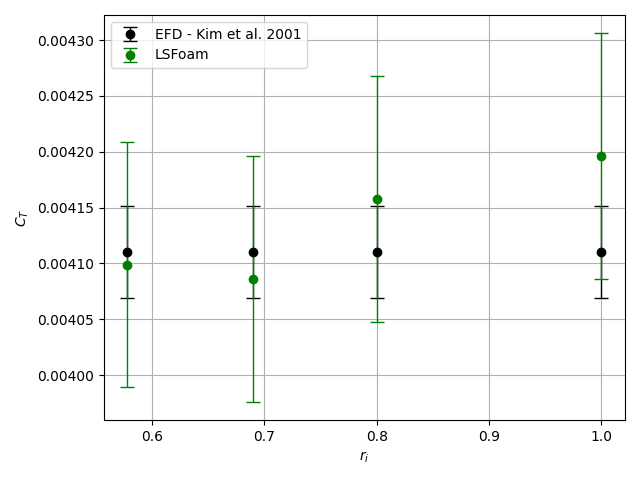}
			\caption{\centering Force coefficient comparison between \textit{LSFoam} and experimental data. The blue bars indicate the experimental uncertainty of 1 \%. The green bars is the mesh uncertainty with a total height of $2U_G$. }
			\label{fig:KVLCC2_ForceCoeff}
		\end{figure}

\begin{figure}[h!]
		\centering
		\includegraphics[scale=0.15]{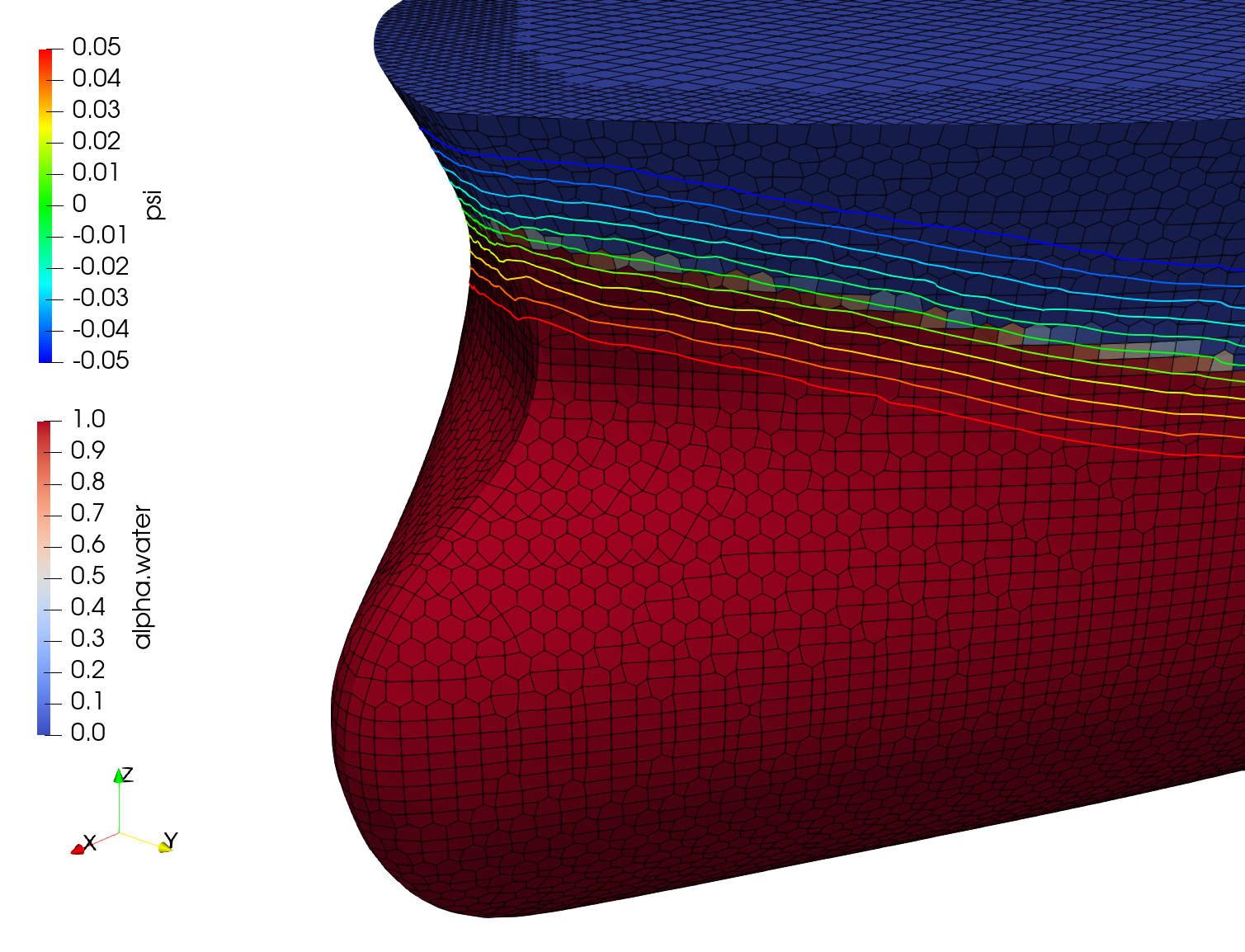}
		\caption{Illustration of hull phase fraction with $\psi(\boldsymbol{x})=$ -0.05 m to 0.05 m each 0.01 m.}
		\label{Hull_alpha}
	\end{figure}

\begin{figure}[h!]
		\centering
		\includegraphics[scale=0.55]{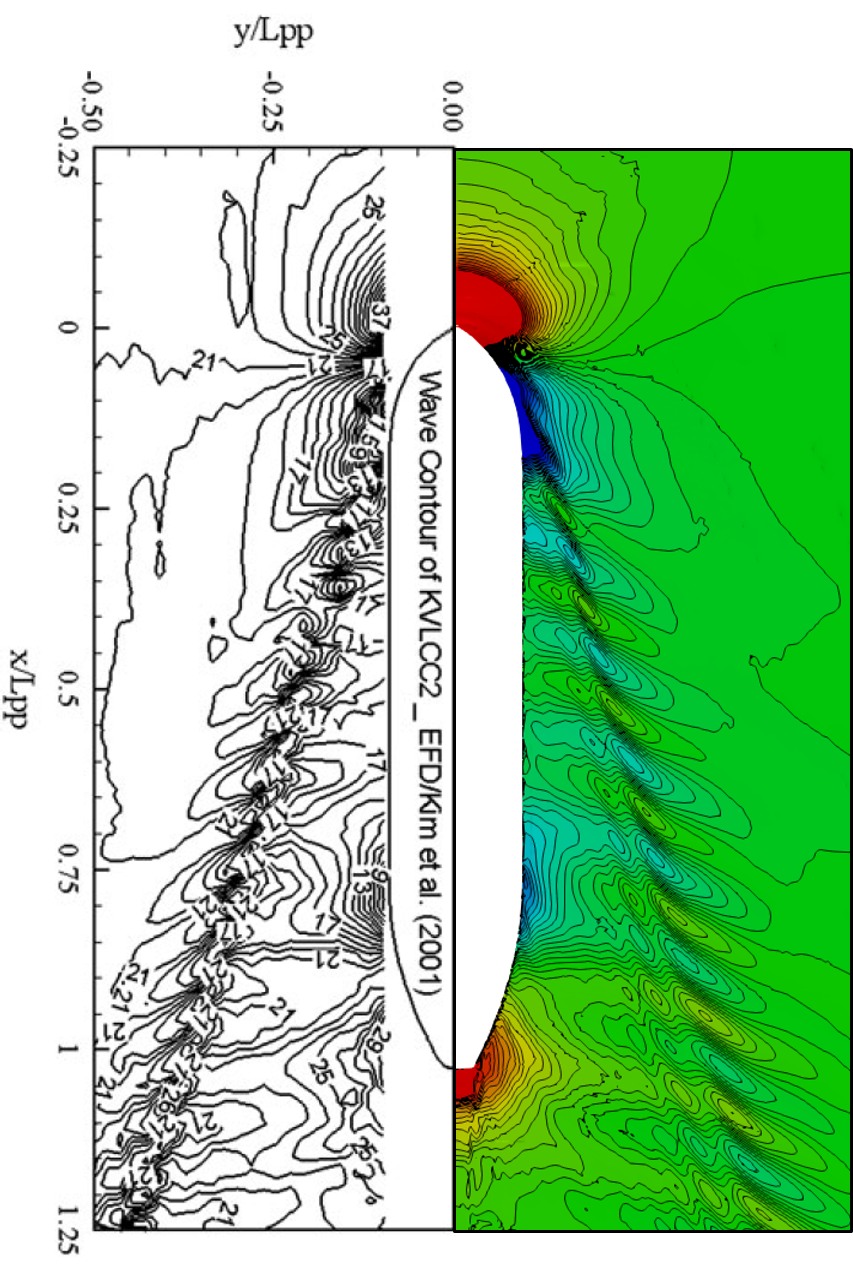}
		\caption{KVLCC2 wave pattern illustration - left EFD - right \textit{LSFoam} finest grid.}
		\label{Wave Patterns}
	\end{figure}

\begin{figure}[h!]
		\centering
		\includegraphics[scale=0.45]{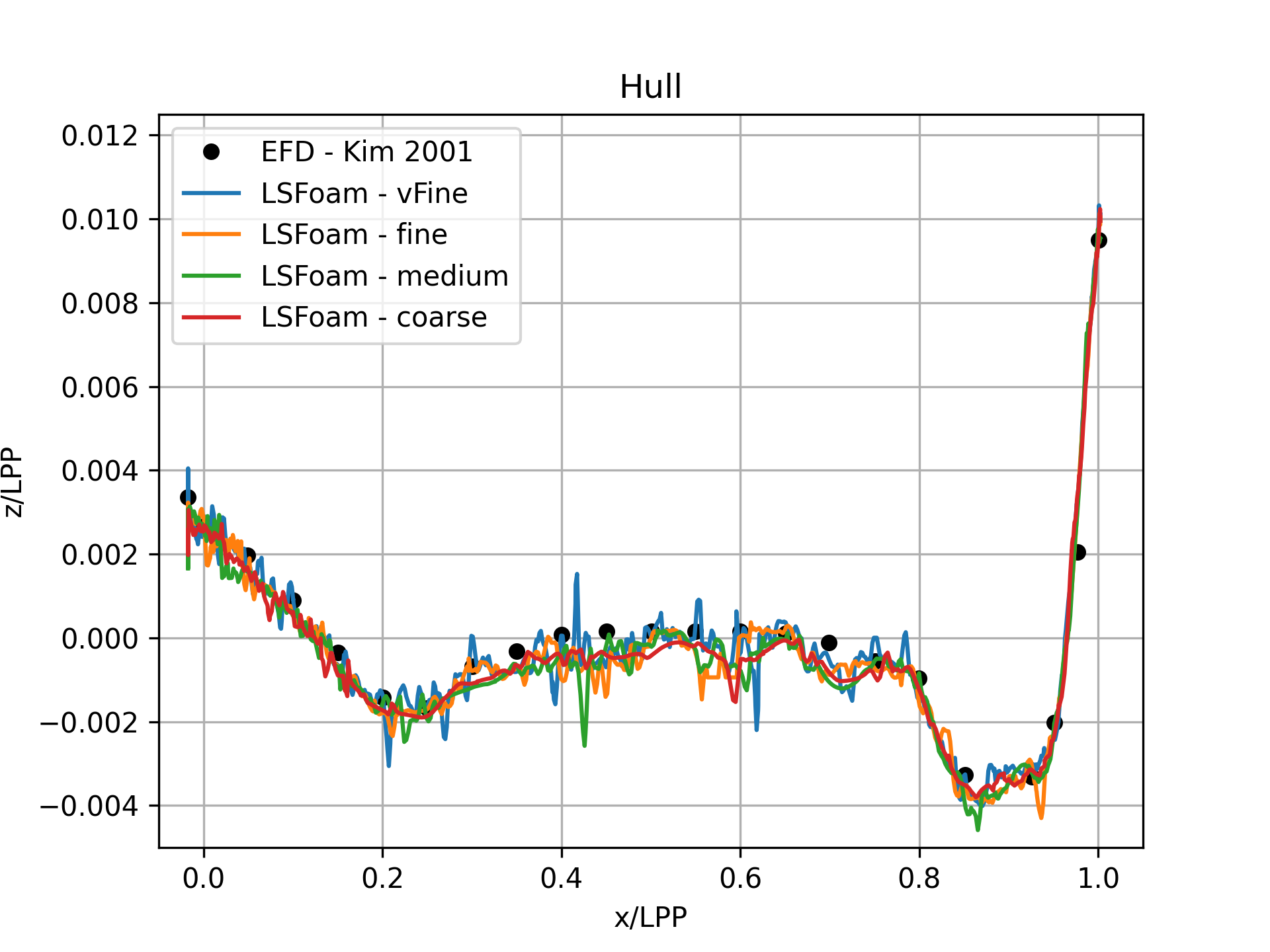}		
		\includegraphics[scale=0.45]{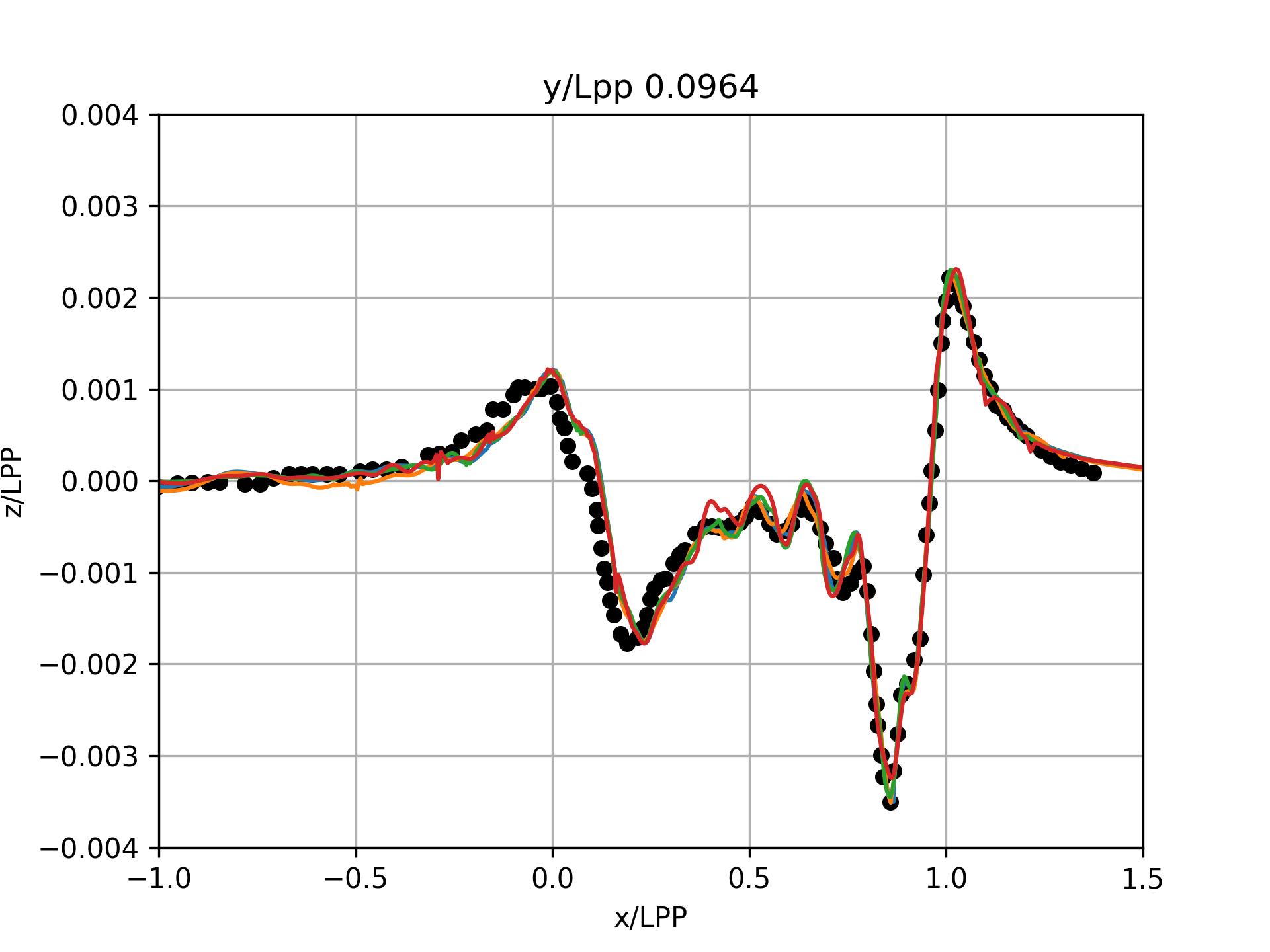}
		\includegraphics[scale=0.45]{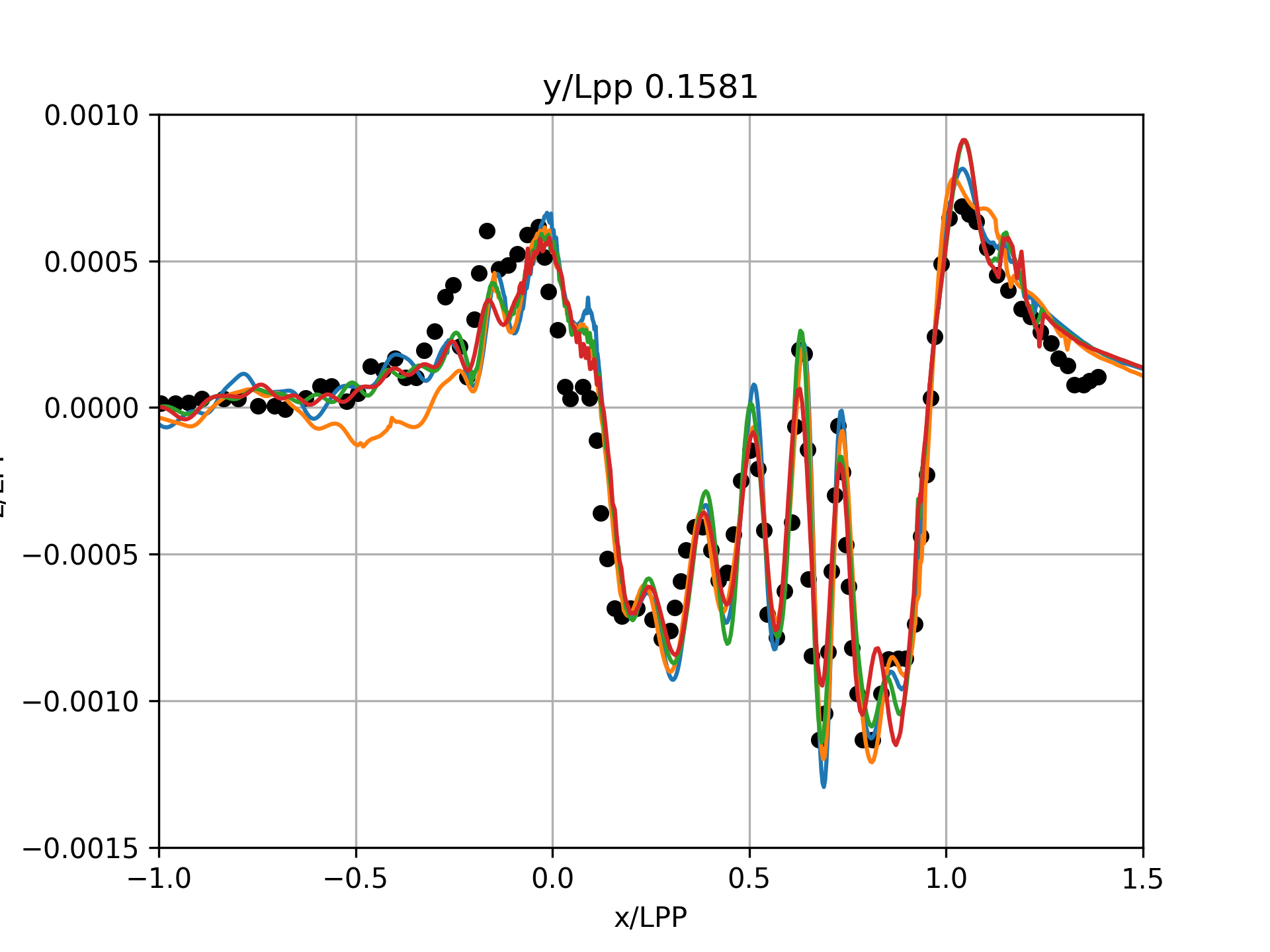}
		\includegraphics[scale=0.45]{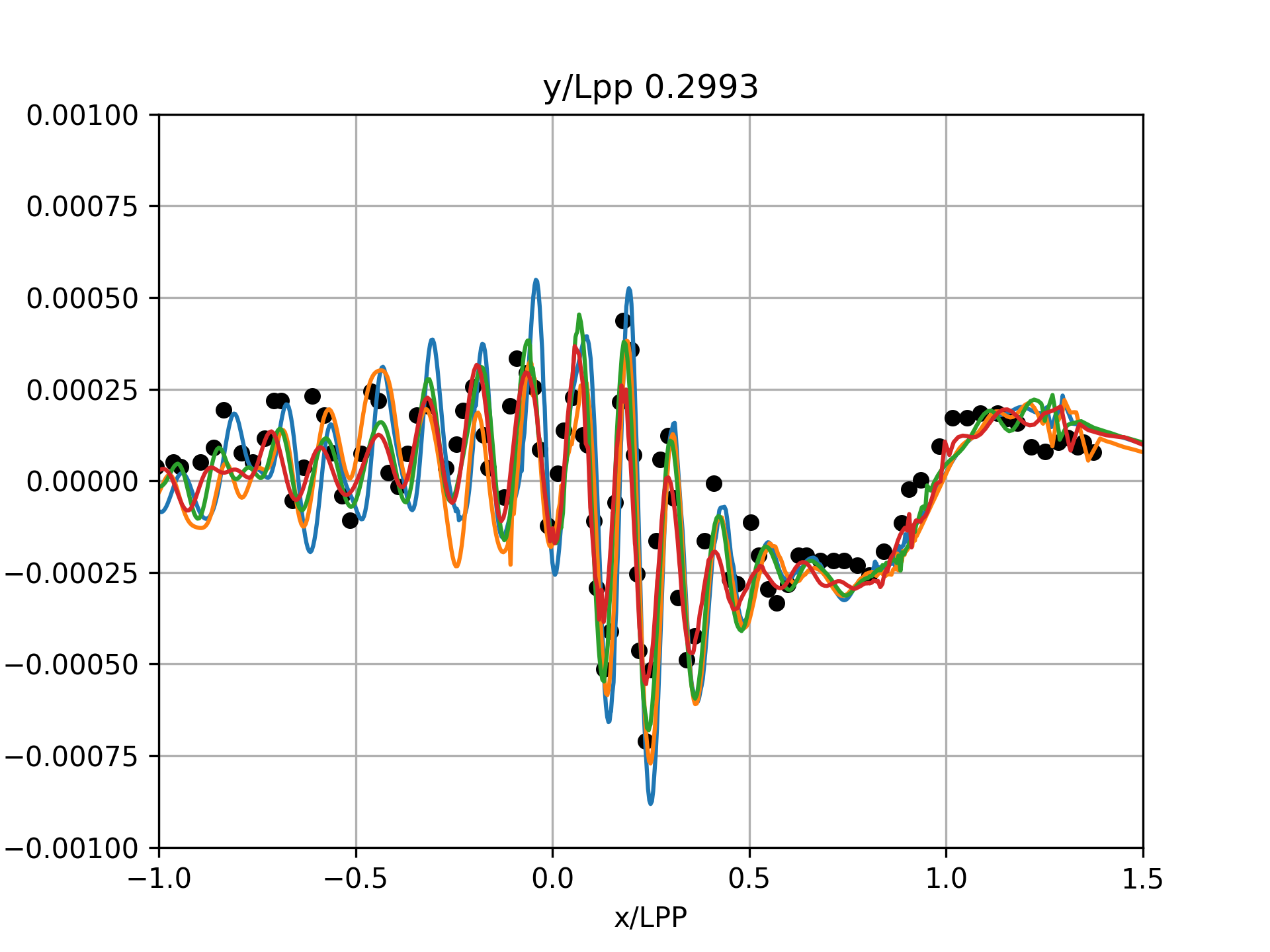}
		\caption{CFD free surface elevation (z/Lpp) comparison with experimental data on y/Lpp lines.}
		\label{Z:LPP KVLCC2}
	\end{figure}

	The KVLCC2 (\citet{Kim2001}) is a model scale ship. Its main particularities
	are detailed in Table \ref{tab:KVLCC2main}. Test cases 1.b and 1.2a
	are parts of the 2010 Gothenburg workshop (\citet{Larsson2014}) and
	consist in simulating a towing tank test with a Froude number equal
	to 0.142 (1.047 m/s). The KVLCC2 test case is challenging for free
	surface methods as the maximum wave height is less than 1\% of $L_{pp}$ \textcolor{black}{(Length Between Perpendicular)}.
	Numerical results are compared with resistance and wave elevation
	measurements. Simulations are carried out in the fixed ship reference
	frame with a symmetry hypothesis on the y = 0 plane. The air/water
	flow is imposed at the entrance of the computational box. A pressure
	reference is imposed on the top through atmospheric boundary conditions.
	For bottom and lateral patches, a slip condition is used, while for
	outlet the ones, zero-gradient conditions are defined. Wall functions are
	used for the hull patch. Four levels of mesh refinement are generated
	with \textit{snappyHexMesh} with respectively 1.7 M, 2.37 M , 3.2 M and 5.5 M cells
	(lately referred to coarse/grid 0, medium/grid 1, fine/grid 2 and
	vfine/grid 3). The average grid refinement ratio between grids, $r_{g}=1.2$,
	is calculated and used for the grid convergence study. The mesh is
	built using several refinement zones, as illustrated on Figure \ref{fig:KVLCCmesh}. \textcolor{black}{The maximum mesh non-orthogonality is 82° and the maximum aspect ratio is 129}.
	In the near-hull region, cells are refined in each direction within
	a rectangular box. Close to the free surface, cells are refined in
	the vertical direction. Moreover, in the Kelvin wake region, cells
	are also refined in longitudinal (Ox) and transversal (Oy) directions.
	Finally, 8 boundary layer cells are inflated with 100 \% of cover layer rate. The time step has been fixed to 20 ms resulting in a maximum Courant number of 50 for the finest grid. The time discretization is achieved with first order Euler scheme because only steady-state results are of interest. The spatial discretization is identical to 3D dambreak simulation and the PIMPLE algorithm stops when the calculated pressure residual is lower than $\epsilon_{p_{d}} < 10^{-5}$. An EARSM turbulence model \cite{Hellsten} is used. The steady state drag coefficients are calculated and compared to experimental
	data for the 3 finest grids, as shown in Figure \ref{fig:KVLCC2_ForceCoeff}.
	Following \citet{Vukcevic2016} guidelines, grid refinement study
	results and validation on grid 1, 2 and 3 are presented in Table \ref{KVLCC2 Errors}. \textcolor{black}{A 6 meter long rectangular} relaxation zone is added \textcolor{black}{near the outlet patch} to damp the free surface and velocity perturbations.
	The resistance coefficient relative errors are lower than 2\%. Figure
	\ref{Hull_alpha} represents the phase fraction and some contours $\psi(\boldsymbol{x})=constant$
	illustrating that the Level-Set function is well preserved. \textit{LSFoam}
	wave patterns, presented on the top side of Figure \ref{Wave Patterns},
	are reasonably matching the experimental data (bottom side of Figure
	\ref{Wave Patterns}). Regarding Figure \ref{Z:LPP KVLCC2}, that
	represents the free surface elevations (z/Lpp for some planes y/Lpp
	= constant), \textit{LSFoam} results fit the experimental data for planes y/Lpp=
	0.09640 and y/Lpp = 0.1581. On the plane y/Lpp = 0.2993, free surface oscillations are slightly overestimated for the finest grid and probably caused by wave reflection due to the mesh transition outside the Kelvin triangle. Regarding mass conservation, for the finest grid, the relative error mass is below $10^{-5}$ showing the ability of the present method to maintain mass and hence the water level.

	\begin{table}[h!]
				\begin{tabular}{|c|c|c|}
					\hline
					Designation & Prototype & TT model\tabularnewline
					\hline
					\hline
					Scale ratio & 1 & 1/58\tabularnewline
					\hline
					Speed (m/s) & 7.9739 & 1.047\tabularnewline
					\hline
					Froude number & 0.142 & 0.142\tabularnewline
					\hline
					Reynolds number & 2.1 x $10^{9}$ & 4.6 x $10^{6}$\tabularnewline
					\hline
					\textcolor{black}{Lpp} (m) & 320 & 5.5172\tabularnewline
					\hline
					Breadth (m) & 58 & 1\tabularnewline
					\hline
					Depth (m) & 30 & 0.5172\tabularnewline
					\hline
					Draft (m) & 20.8 & 0.3586\tabularnewline
					\hline
					Wetted surface area (m\texttwosuperior ) & 27194 & 8038.8\tabularnewline
					\hline
					Displacement (m3) & 312.621 & 1.6023\tabularnewline
					\hline
					Block coefficient & 0.8098 & 0.8098\tabularnewline
					\hline
				\end{tabular}
				\centering
				\caption{KVLCC2 main particulars. }
				\label{tab:KVLCC2main}
		\end{table}

		\newpage
		
		\begin{table}[h!]
			\centering
			\begin{tabular}{|c|c|}
				\hline
				Measurement Result & $C_{T}$\tabularnewline
				\hline
				\hline
				Experimental Result & 0.00411\tabularnewline
				\hline
				Experimental uncertainty (\%) & 1 \%\tabularnewline
				\hline
				Grid 1 solution & 0.004158\tabularnewline
				\hline
				Grid 2 solution & 0.004086\tabularnewline
				\hline
				Grid 3 solution & 0.004099\tabularnewline
				\hline
				Error grid 1 & 0.00004\tabularnewline
				\hline
				Error grid 2 & -0.00002\tabularnewline
				\hline
				Error grid 3 & -0.00001\tabularnewline
				\hline
				Relative error grid 1 (\%) & 1.16 \%\tabularnewline
				\hline
				Relative error grid 2 (\%) & -0.58 \%\tabularnewline
				\hline
				Relative error grid 3 (\%) & -0.26 \%\tabularnewline
				\hline
				$R_{i}$  &  $-1 < R_{i} <0$, oscillatory convergence \tabularnewline
				\hline
				$U_{SN}$  & 0.01 \%\tabularnewline
				\hline
				$U_{G}$ & 1.3 \%\tabularnewline
				\hline
				$U_{v}$  & 0.01 \%\tabularnewline
				\hline
			\end{tabular}\caption{KVLCC2 validation and grid refinement study for the three finest grid. \label{KVLCC2 Errors}}
		\end{table}

	\section{Conclusions}

	The Level-Set algebraic-like approach (\citet{SUSSMAN1994146}) is
	based on a function $\psi$ being the signed distance to the interface.
	To capture interface motions, this function is advected in a flow
	field. Such a procedure will break its distance property and lead
	to unacceptable mass variations. To tackle these issues, the reinitialization equation \ref{eq:-1}
	from \citet{SUSSMAN1994146} is traditionally solved explicitly \textcolor{black}{with a uniform spatial time step}, but many practical limits remain: robustness, reinitialization frequency, number of iterations, mass variation, and polyhedral meshes \textcolor{black}{with industrial quality (mainly high non-orthogonality and aspect ratio)}. In this work, we propose to address all of these issues by:
	\begin{itemize}
		\item Adopting an implicit form of the reinitialization equation with LTS time advancement,
		\item Implementing an adaptive thickness size for sign and filtering functions,
		\item Enforcing the immobility of the interface during reinitialization iteration through the marking of anchoring cells.
	\end{itemize}
	
     \textcolor{black}{The benefits of the LTS approach and anchoring cells are demonstrated in chapter \ref{LS_testcase1}}. For pressure-velocity coupling, in a perspective to enhanced both accuracy and robustness, the solver takes advantage of the PIMPLE algorithm with a consistent momentum interpolation formulation \cite{Cubero2007} and the GFM \cite{Fedwick1999} to handle density discontinuities \cite{Vukcevic2017}. With this Level-Set approach, the GFM is eased by the direct calculation of the distance to the interface.  
	
	The present approach has been coded in the OpenFOAM (\cite{Weller1998}) framework within a new solver, named \textit{LSFoam}, and has been tested on five test cases covering different flow configurations: the rising bubble test case, \citet{Hysing2007}, Rayleigh-Taylor instability simulations (\citet{Puckett1997}), Ogee spillway flow \cite{erpicum2018experimental}, tridimensional dambreak simulation with a square cylinder obstacle \citet{GomezGesteira2013} and KVLCC2 steady resistance calculations (\citet{Larsson2014}). For the first two cases with surface tension dominated flow, the solver gave results very close to reference solutions. To challenge the method, non-uniform and unstructured grids have been used for the last three cases. For the Ogee spillway test, the method efficiently maintains the water level. The overall results are in medium agreement with the data of \cite{erpicum2018experimental}, although the flow detachment is well predicted. Deviations in the discharge coefficient are observed, but are likely caused by turbulence modeling and/or the 2D assumption. For the dambreak simulation, the results are in good agreement with the experimental data. For ship resistance applications, the present method has shown excellent mass conservation properties as well as force calculations and wave patterns. Regarding mass losses, the conservation is excellent for all cases excepting for the 3D dambreak, where the motion of the free surface is very complex, although the mass error is limited to 1\%. A potential solution could be to implement a method to enforce mass conservations for such cases. \textcolor{black}{Gathering all the numerics detailed in this work, \textit{LSFoam} has shown excellent robustness by handling mesh non-orthogonality above 80° and maximal Courant number higher than 100}. The original approach of \citet{SUSSMAN1994146} with adequate enhancement is able to give overall excellent results for different flow configurations. The method is particularly promising for moderate free surface deformations typically encountered in marine and offshore applications.

	\section{Declaration of competing interests}
	The authors declare that they have no known competing financial.

\bibliography{references}
\end{document}